\documentclass{aastex}
\usepackage{emulateapj5}
\usepackage{epsf}
\usepackage{onecolfloat}

\newcommand{\rr}{\mbox{\bf r}}
\newcommand{\RR}{\mbox{\bf R}}

\begin{document}

%\submitted{submitted to ApJ}

\twocolumn[

\title{The tumultuous lives of galactic dwarfs\\
  and the missing satellites problem}

\author{Andrey V. Kravtsov\altaffilmark{1}, Oleg Y. Gnedin\altaffilmark{2}, 
Anatoly A. Klypin\altaffilmark{3}}
%]
%\twocolumn[

\begin{abstract}  
  Hierarchical Cold Dark Matter (CDM) models predict that Milky Way
  sized halos contain several hundred dense low-mass dark matter
  satellites (the substructure), an order of magnitude more than the
  number of observed satellites in the Local Group. If the CDM
  paradigm is correct, this prediction implies that the Milky Way and
  Andromeda are filled with numerous dark halos. To understand why
  these halos failed to form stars and become galaxies, we need to
  understand their history.  We analyze the dynamical evolution of the
  substructure halos in a high-resolution cosmological simulation of
  Milky Way sized halos in the $\Lambda$CDM cosmology. We find that
  about 10\% of the substructure halos with the present masses
  $\lesssim 10^8-10^9\rm\ M_{\odot}$ (circular velocities $V_{\rm
    m}\lesssim 30\rm\ km/s)$ had considerably larger masses and
  circular velocities when they formed at redshifts $z\gtrsim 2$.
  After the initial period of mass accretion in isolation, these
  objects experience dramatic mass loss due to tidal stripping.  Our
  analysis shows that strong tidal interaction is often caused by
  actively merging massive neighboring halos, even before the
  satellites are accreted by their host halo. These results can
  explain how the smallest dwarf spheroidal galaxies of the Local
  Group were able to build up a sizable stellar mass in their
  seemingly shallow potential wells.  We propose a new model in which
  all of the luminous dwarf spheroidals in the Local Group are
  descendants of the relatively massive ($\gtrsim 10^9\rm \ 
  M_{\odot}$) high-redshift systems, in which the gas could cool
  efficiently by atomic line emission and which were not significantly
  affected by the extragalactic ultraviolet radiation.  We present a
  simple galaxy formation model based on the trajectories extracted
  from the simulation, which accounts for the bursts of star formation
  after strong tidal shocks and the inefficiency of gas cooling in
  halos with virial temperatures $T_{\rm vir} \lesssim 10^4$~K.  Our
  model reproduces the abundance, spatial distribution, and
  morphological segregation of the observed Galactic satellites. The
  results are insensitive to the redshift of reionization.
\end{abstract}

% User-supplied List of keywords.

 ]

\altaffiltext{1}{Dept. of Astronomy and Astrophysics,
       Center for Cosmological Physics,
       The University of Chicago, Chicago, IL 60637;
       {\tt andrey@oddjob.uchicago.edu}}
\altaffiltext{2}{Space Telescope Science Institute,
       3700 San Martin Drive, Baltimore, MD 21218;
       {\tt ognedin@stsci.edu}}
\altaffiltext{3}{Astronomy Department, New Mexico State University,
MSC 4500, P.O.Box 30001, Las Cruces, NM, 880003-8001;
       {\tt aklypin@nmsu.edu}}

%----------------------
\section{Introduction}
\label{sec:intro}
%----------------------

Semi-analytic models of galaxy formation
\citep{kauffman_etal93,bullock_etal00,somerville02,benson_etal02} and
numerical simulations \citep{klypin_etal99a,moore_etal99} have convincingly
showed that the expected number of dark matter clumps around the
galactic Milky Way (MW) sized halos exceeds the observed number of
satellites by an order of magnitude. The discrepancy may indicate that
the amplitude of the small-scale primordial density fluctuations is
considerably lower than expected in the Cold Dark Matter (CDM)
scenarios \citep[e.g.,][]{kamionkowski_liddle00,zentner_bullock03} or
that dark matter is self-interacting \citep{spergel_steinhardt00}. An
alternative ``astrophysical'' interpretation is that the mismatch
indicates that galaxy formation in dwarf halos is inefficient.

Several plausible physical processes may suppress gas accretion and
star formation in dwarf dark matter (DM) halos.  The cosmological UV
background, which reionized the Universe at $z> 6$, heats the
intergalactic gas and establishes a characteristic time-dependent
minimum mass for halos that can accrete gas
\citep[e.g.,][]{efstathiou92,thoul_weinberg96,quinn_etal96,navarro_steinmetz97,gnedin_hui98,kitayama_ikeuchi00,gnedin00,dijkstra_etal03}.
The gas in the low-mass halos may be photoevaporated after reionization
% even if they accrete some gas early on
\citep{barkana_loeb99,shaviv_dekel03,shapiro_etal03}. In particular,
\citet{shaviv_dekel03} recently argued that halos with circular
velocities of up to $\sim 30\ \rm km\,s^{-1}$ can be photo-evaporated
by the UV background. At the same time, the ionizing radiation may quickly
dissociate molecular hydrogen, the only efficient coolant for
low-metallicity gas in such halos, and prevent star formation before
the gas is completely removed \citep[e.g.,][]{haiman_etal97}.

The combined effect of these processes is likely to leave all DM halos
with masses $\lesssim {\rm few}\times 10^9\ \rm M_{\odot}$ dark.  This
is consistent with current observational constraints which indicate
that halos with $M<10^{10}\ \rm M_{\odot}$ are virtually devoid of
galaxies \citep{vandenbosch_etal03}. It is thus remarkable that the
dynamical masses of some of the Local Group dwarfs are only $\sim
10^7\ \rm M_{\odot}$ \citep{mateo98}. How could such galaxies form stars
despite the suppressing processes listed above? 

One possibility is that they manage to accrete a certain amount of gas
before the Universe is reionized \citep{bullock_etal00} with the
implicit assumption that this gas can be subsequently converted to
stars. However, it is likely that gas cooling and star formation in
such small systems is inefficient. For example, cosmological
simulations with self-consistent treatment of H$_2$ chemistry and
radiative transfer indicate that star formation is strongly suppressed
in halos with masses $M\lesssim 5\times 10^8\ \rm M_{\odot}$ at all
redshifts, even before reionization \citep{chiu_etal01}. In addition,
the galaxies may not be able to form sufficiently early to accrete the
gas in the first place, if the power spectrum normalization is low or
the Universe was reionized early, as indicated by the first-year {\sl
  WMAP} results \citep{spergel_etal03,kogut_etal03}. An alternative
proposal was recently suggested by \citet{stoehr_etal02,stoehr_etal03}
and corraborated by \citet{hayashi_etal03}, who argued that 
the host halos of the low-luminosity dwarf spheroidal galaxies 
may be considerably more massive than previously thought. In this case,
the large halo mass could allow an object to resist the suppressing
effects of UV background.

In this paper we study the dynamical evolution of dwarf satellite halos
around the Milky Way-sized hosts in self-consistent cosmological
simulations.  We show that the evolution of such objects is
complex and often involves dramatic tidal stripping, interactions
with other satellites, mass loss, and changes of internal structure.
Most importantly, we find that some of the satellites that have small
masses and circular velocities at the present, once were
considerably more massive and could have plausibly formed stars in the past.
We argue that the evolution of these objects may explain how the
smallest dwarfs in
the Local Group managed to form their stellar populations.

The paper is organized as follows. In \S~\ref{sec:sim} we describe
the details of the numerical simulation used in our
analysis. In \S~\ref{sec:haloid} and \S~\ref{sec:traj} we discuss the
algorithm used to identify halos and the method used to construct
their evolutionary tracks.  In
\S~\ref{sec:results} we present the main results on the dynamical
evolution, abundance, and radial distribution of the dwarf dark
matter halos. In \S~\ref{sec:sam} we present a model
for star formation in these systems and compare results to the observed
abundance and spatial distribution of the Local Group dwarfs. 
We discuss the implications of our results and compare our model to the 
previous studies in \S\S~\ref{sec:discussion} and \ref{sec:prevwork}.
Finally, in \S~\ref{sec:conclusions} we summarize our findings and 
conclusions.

%----------------------
\section{Simulation}
\label{sec:sim}
%----------------------

We used the Adaptive Refinement Tree $N$-body code \citep[ART,
][]{kravtsov_etal97,kravtsov99} to follow the evolution of three
Milky Way-sized halos in the concordance $\Lambda$CDM cosmology:
$(\Omega_{\rm m}, \Omega_{\Lambda}, h, \sigma_8)=(0.3, 0.7, 0.7,
0.9)$.  The simulation starts with a uniform $256^3$ grid covering the
entire computational box.  This grid defines the lowest (zeroth) level
of resolution.  Higher force resolution is achieved in the regions
corresponding to collapsing structures by recursive refining of all
such regions using an adaptive refinement algorithm. Each cell can be
refined or de-refined individually. The cells are refined if the particle
mass contained within them exceeds a certain specified value. The grid
is thus refined to follow the collapsing objects in a quasi-lagrangian
fashion.

The galactic halos were simulated in the comoving box of $25h^{-1}$ Mpc; they
were selected to reside in a well-defined filament at $z=0$.  Two
halos are neighbors, located at $425h^{-1}\ \rm kpc$ (i.e., $\approx 610\ 
{\rm kpc}\sim 2R_{\rm vir}$) from each other. The configuration of
this pair thus resembles that of the Local Group. The third halo is
isolated and is located $\sim 2$~Mpc away from the pair.

Multiple mass resolution technique was used to set up initial
conditions.  Namely, a lagrangian region corresponding to a sphere of
radius equal to two virial radii around each halo was re-sampled with
the highest resolution particles of mass $m_{\rm p}=1.2\times
10^6h^{-1}{\rm M_{\odot}}$, corresponding to $1024^3$ particles in the
box, at the initial redshift of the simulation ($z_{\rm i}=50$). The
high mass resolution region was surrounded by layers of particles of
increasing mass with a total of 5 particle species. Only regions
containing highest resolution particles were adaptively refined and
the threshold for refinement was set to correspond to the mass of the four
highest resolution particles. The maximum of ten refinement levels
was reached in the simulations corresponding to the peak formal
spatial resolution of $150$ comoving parsec.  Each host halo is
resolved with $\sim 10^6$ particles within its virial radius at $z=0$.

From this point, we will refer to the isolated halo as G$_1$ and the
halos in pair as G$_2$ and G$_3$. These halos are called B$_1$, C$_1$,
and D$_1$, respectively, in \citet{klypin_etal01} and we refer the
reader to this paper for further details. The main properties of these
three host halos, the virial mass, radius, and maximum circular
velocity, are given in Table~1. We choose to define the virial radius
(and the corresponding virial mass) as the radius encompassing the
density of 180 times the mean density of the universe. For the
commonly used overdensity of 340, the virial radii and masses for
G$_1$, G$_2$, and G$_3$ are $R_{340}=231$, $212$, and $213h^{-1}$~kpc
and $M_{340}=1.45\times 10^{12}$, $1.13\times 10^{12}$, and
$1.14\times 10^{12}h^{-1}\ \rm M_{\odot}$, respectively. The masses
are in the range of possible MW halo masses \citep{klypin_etal02}.

\begin{table}[t]
\label{tab:hosts}
\begin{center}
{\sc Table 1. Properties of the host halos} \\[2mm]
\begin{tabular}{ccccc}
\tableline\tableline\\
\multicolumn{1}{c}{Halo} &
\multicolumn{1}{c}{$M_{\rm vir}$} &
\multicolumn{1}{c}{$R_{\rm vir}$} &
\multicolumn{1}{c}{$V_{\rm m}$} &
\multicolumn{1}{c}{Environment}
\\
& \multicolumn{1}{c}{($h^{-1}{\rm M_{\odot}}$)} & 
\multicolumn{1}{c}{($h^{-1}$~kpc)} &
\multicolumn{1}{c}{($\rm km\,s^{-1}$)} &
\\[2mm]
\tableline\\
G$_1$ & $1.66\times 10^{12}$ & 298  & $213$ & isolated\\ 
G$_2$ & $1.24\times 10^{12}$ & 278  & $199$ & pair \\
G$_3$ & $1.19\times 10^{12}$ & 281  & $183$ & pair \\ 
% G1 -> H4, G2-> H1, G3-> H2
\\
\tableline
\end{tabular}
\end{center}
{\small Note -- $R_{\rm vir}$ is the virial radius corresponding to the
average density of $180$ times the mean density of the universe in $h^{-1}$~kpc; $M_{\rm vir}=M(<R_{\rm vir})$ in $h^{-1}{\rm M_{\odot}}$
(both radius and mass are given for $z=0$); $V_{\rm m}$ is the maximum
circular velocity.}
\end{table}

\begin{figure}[t]
\vspace{-0.5cm}
\centerline{\epsfysize4truein \epsffile{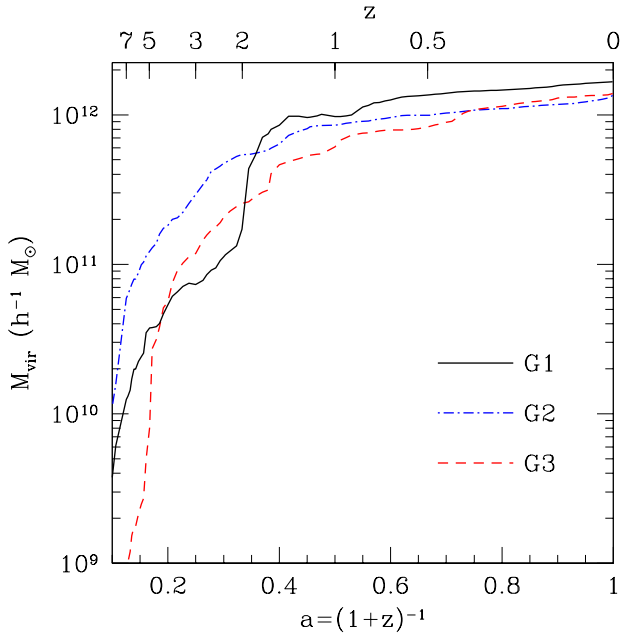}}
\vspace{-1cm}
\caption{Mass aggregation histories for the three MW-size host halos 
analyzed in this study. 
\label{fig:hmah}}
\end{figure}

Figure~\ref{fig:hmah} shows the mass aggregation history of the three
host halos. They have similar masses at the present
but rather different evolutionary histories. In all cases, 
there is a period of very rapid mass assembly at $z\gtrsim 2-3$
followed by a relatively quiescent accretion at $z\lesssim 1.5$, the behavior
typical of hierarchically forming halos \citep{wechsler_etal02}. Host
G$_1$ undergoes a spectacular multiple major merger at $z\approx 2$,
which results in a dramatic mass increase on a dynamical time scale.
Halos G$_2$ and G$_3$ increase their mass in a series of somewhat less
spectacular major mergers which could be seen as mass jumps at $5 < z < 1$.
All three systems accrete little mass and experience no major
mergers at $z\lesssim 1$ (or lookback time of $\approx 8$~Gyr) and
thus could host a disk galaxy. Note, however, that halos
G$_1$ and G$_3$ experience minor mergers during this period.

%=======================
\section{Halo identification}
\label{sec:haloid}
%=======================

\begin{figure}[t]
\plotone{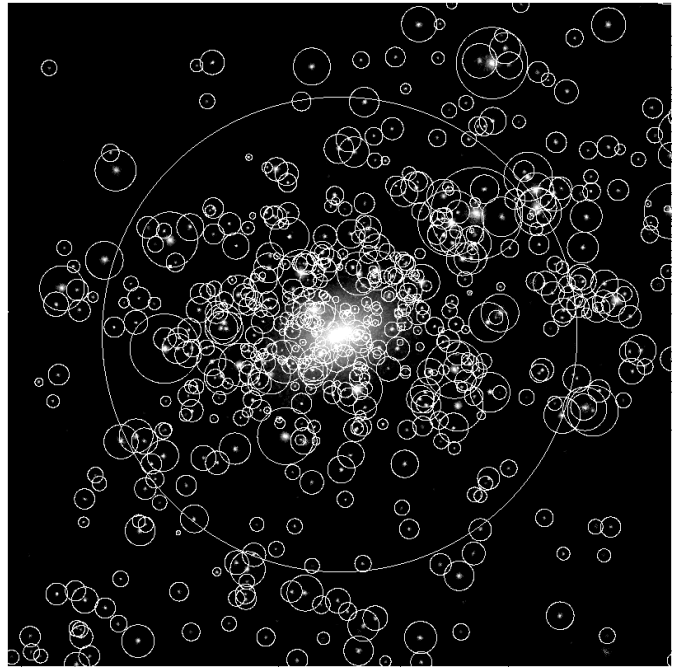}
\caption{Distribution of dark matter particles (points) and dark matter halos
  (circles) identified by our halo finding algorithm centered on the
  isolated galactic halo at $z=0$.  The radius of the largest circle
  indicates the actual virial radius, $R_{\rm vir}$, of the host halo
  ($R_{\rm vir}=298h^{-1}$~kpc); the radii of the other halos are the
  minimum between truncation radius $r_{\rm t}$ and $R_{\rm vir}$. The
  particles are colored on a gray-scale logarithmic stretch according
  to their local density. The stretch is chosen to highlight the cores of 
the halos for clarity.  
  \label{fig:dmh}}
\end{figure}

In this study we use a variant of the Bound Density Maxima \citep[BDM,][]{klypin_etal99} halo
finding algorithm to identify halos both within (subhalos) and outside
the host halos. Throughout this paper we will use the terms {\em subhalo},
{\em substructure}, and {\em satellite} interchangeably to indicate the 
distinct gravitationally self-bound halos located within the virial
radius of a larger halo, which we call the {\em host}. The division
is illustrated in Figure~\ref{fig:dmh}.

The BDM algorithm first finds positions of local maxima in the density field
smoothed on a certain scale.  Starting with the highest overdensity
particle, we surround each potential density maximum by a sphere of radius
$r_{\rm find}=10h^{-1}\ \rm kpc$ and exclude all particles within this
sphere from further search. The search radius is defined by the size
of smallest systems we aim to identify. We verified that the results
do not change if this radius is decreased by a factor of up to four.
After all potential halo centers are identified, we analyze the density
distribution and velocities of surrounding particles to test whether
the center corresponds to a gravitationally bound clump. Specifically,
we construct the density, circular velocity, and velocity dispersion
profiles around each center and iteratively remove unbound particles
 \citep[see][for details]{klypin_etal99}. We then
construct final profiles using only bound particles and use them to
calculate such halo properties as the maximum circular velocity
$V_{\rm m}$, mass $M$, etc.

The virial radius is meaningless for the subhalos within a larger host
as their outer layers are tidally stripped and the extent of the halo
is truncated. The definition of the outer boundary of a subhalo and
its mass are thus somewhat ambiguous.  We adopt the {\rm truncation
radius}, $r_{\rm t}$, at which the logarithmic slope of the density
profile constructed from the bound particles becomes larger than
$-0.5$ as we do not expect the density profile of the CDM halos to be
flatter than this slope. Empirically, this definition roughly
corresponds to the radius at which the density of the gravitationally
bound particles is equal to the background host halo density, albeit
with a large scatter. For some halos $r_{\rm t}$ is larger than their
virial radius. In this case, we set $r_{\rm t}=R_{\rm vir}$.
Throughout this paper, we will denote the minimum of the virial mass
and mass within $r_{\rm t}$, simply as $M$. For each halo we also
construct the circular velocity profile $V_{c}(r)=\sqrt{GM(<r)/r}$ and
compute the maximum circular velocity profile $V_{\rm m}$.

Figure~\ref{fig:dmh} shows the particle distribution in the halo G$_1$
at $z=0$ along with the halos (circles) identified by the halo finder.
The particles are color-coded on a gray scale according to the
logarithm of their density to enhance visibility of substructure
clumps. The radius of the largest circle indicates the actual virial
radius, $R_{\rm vir}$, of the host halo ($R_{\rm vir}=298h^{-1}$~kpc);
the radii of the other halos are the minimum of the truncation radius
$r_{\rm t}$ and $R_{\rm vir}$. The figure demonstrates that the algorithm is
efficient in identifying the substructure down to small masses.

%--------------------------------------
\section{Constructing trajectories}
\label{sec:traj}
%--------------------------------------

The halo finder described above was run at the 96 saved epochs between
$z=10$ and $z=0$ with a typical spacing of $\sim 1-2\times 10^8$~yr
between outputs. For each epoch, the halo finder produced a halo
catalog with positions, velocities, radii $r_{\rm h}=\min (r_{\rm
  t},r_{\rm vir})$, masses $m(<r_{\rm h})$, maximum of the circular
velocity profile $V_{\rm m}$ and the radius at which the maximum
occurs $r_{\rm max}$. In addition, for each halo we save indices of
all gravitationally-bound DM particles located within $r_{\rm h}$.

This information is used to identify the progenitors of halos at
successive epochs. Specifically, for a current epoch $z_{\rm i}$,
starting at $z=0$, we search progenitors for each halo at several
previous epochs $z_{\rm i-j}$ as follows. First, we select a given
fraction, $f_{\rm bound}$, of the most bound particles of the halos at
the epochs of consideration. We then compare the fraction of these
particles that is common between all pairs of halos at successive
epochs and assume that the halo with the highest common fraction is
the progenitor. The trajectories used in this study were constructed
using $f_{\rm bound}=0.25$.  As the halo catalogs may miss some halos,
especially near the completeness limit of the simulation, if
the progenitor is not found at the previous epoch we need to search at the
earlier epoch, etc. In particular, if a halo is located within the
search radius $r_{\rm search}$ of some larger system it will not be
identified by the halo finder. In constructing the trajectories we
search for progenitors of a halo at $z_{\rm i}$ at epochs up to
$z_{\rm i-4}$. In the dominant majority of cases the progenitors are
found at the previous epoch $z_{\rm i-1}$.  We experimented with other
algorithms for progenitor identification and found the adopted
prescription to be the most reliable and efficient.

\begin{figure*}[tp]
\vspace{-6.5cm}
\centerline{\epsfxsize=\textwidth\epsffile{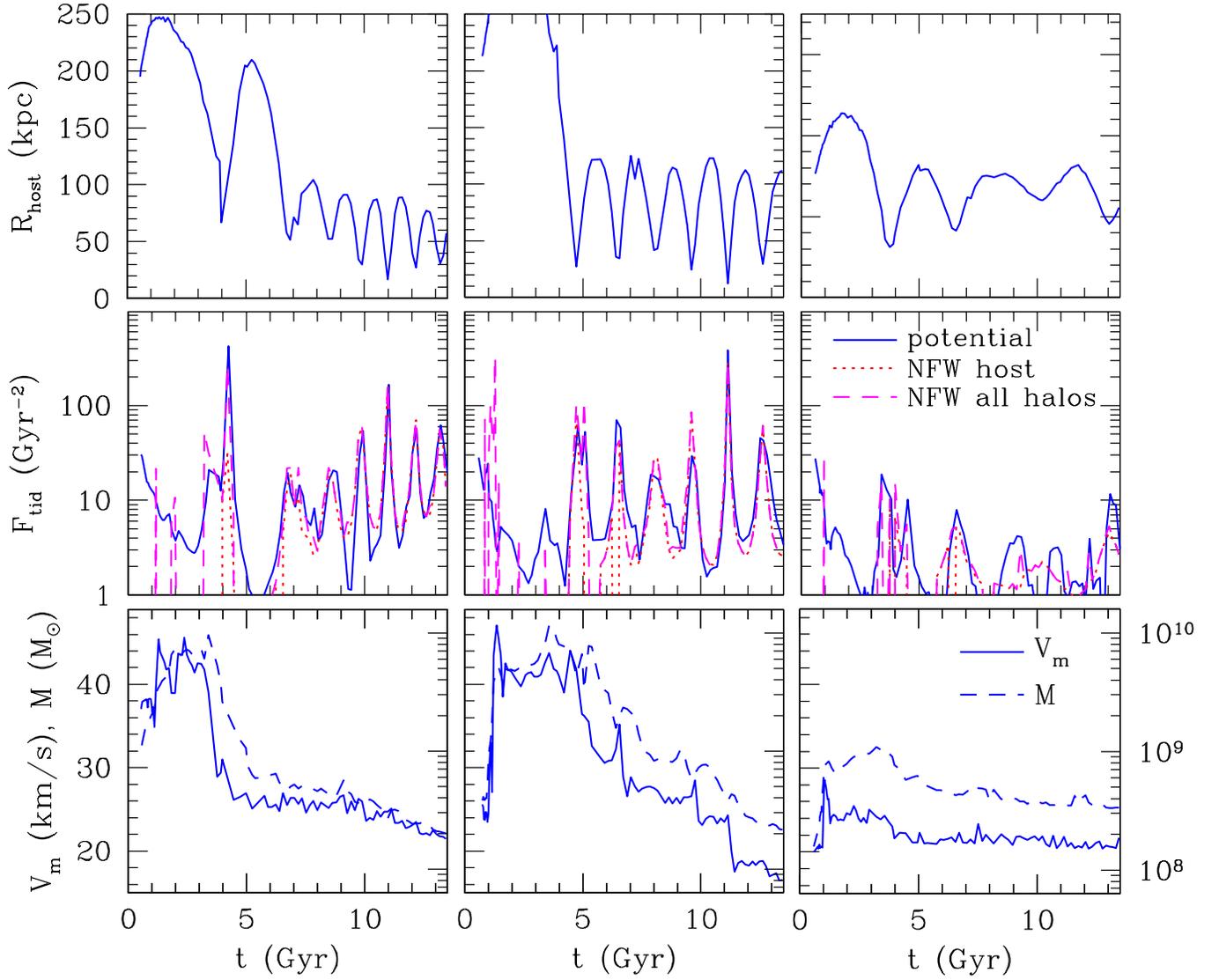}\vspace{-1.75cm}}
\caption{Three examples  
  of the evolution of satellites of a MW-size host (different columns).
  {\it Top panels:} the proper distance between the satellite and the
  center of the host halo as a function of time.  {\it Middle panels:}
  the tidal force experienced by the object, calculated directly from
  gravitational potential, is shown by the {\it solid line\/}.  {\it
    Dotted line} shows the equivalent tidal force from the host halo
  with the host density profile approximated by an NFW model.  {\it
    Dashed line} shows the contributions of all neighboring halos,
  including the host, with the density profiles of halos approximated
  by an NFW model with $r_{\rm max}$ and $V_{\rm m}$ as measured in
  the simulation. See Appendix~\ref{sec:tidalmodel} for details on the
  tidal force calculation.  {\it Bottom panels:} maximum circular
  velocity $V_{\rm m}$ ({\it solid lines}) and bound mass $m(<r_{\rm
    t})$ ({\it dashed lines}) as a function of time.  The three
  objects show different types of evolution: dramatic early stripping
  with a relatively quiescent evolution afterward (left), continuous
  dramatic tidal stripping (middle), weak stripping and quiescent
  evolution (right).
 \label{fig:tr1}}
\end{figure*}

%----------------
\section{Results}
\label{sec:results}
%----------------

%------------------------------------------------
\subsection{Tidal stripping and dynamical evolution of satellite halos}
\label{sec:traj_res}
%------------------------------------------------

Figure~\ref{fig:tr1} shows three examples of the evolution of
satellite halos.  In the middle row of panels we plot the tidal force
experienced by each object. The force was calculated both directly
from the gravitational potential field computed in the simulation and
analytically from the neighbor halo catalogs, as described in
Appendix~\ref{sec:tidalmodel}.  The Figure shows the trace of the
tidal tensor, $F_{\rm tid} = \sum_\alpha F_{\alpha\alpha}$, which is a
good measure of the overall tidal field.

In all cases the tidal force experienced by the satellite coincides
approximately with the time when the object is closest to the host, as
expected.  The Figure shows, for example, that at later epochs the
tidal force calculated directly using the potential from the
simulation can be well approximated by the analytical force from the
host halo (see eq.~\ref{eq:Ftidnfw}).  However, at earlier epochs
(e.g., the highest peak in the left column) the force from the host
underestimates the total tidal force.  Thus, the overall tidal
stripping is produced not only by the host halo but also by the
massive neighbor halos, even {\it before the host is formed}
\citep{gnedin03a}.

The tidal heating by multiple halos is similar to ``galaxy
harassment'' in clusters of galaxies
\citep{moore_etal96,moore_etal99b}, except that it may occur when the
halo is still isolated.  As Figure~\ref{fig:tr1} shows, the true force
computed from the potential can be recovered if the analytical
contributions of all neighboring halos are included.  Their
contribution is particularly important during major mergers of the
host, when the centers of two or more massive halos are located in the
close vicinity of each other and the satellite halos.  Our analytic
estimate describes the strong tidal peaks remarkably well, but becomes
inaccurate for low (a few Gyr$^{-2}$) values of $F_{\rm tid}$.

The amount of energy imparted to the halo depends on the square of the
tidal force (eq.~\ref{eq:Itid}).  Thus, by far the strongest tidal
heating experienced by an object is during the highest tidal peaks.
For the object in the left column of Figure \ref{fig:tr1}, for
example, most of the stripping and disruption is due to the tidal peak
at $t\sim 4$~Gyr ($z\sim 1.5$).  At this epoch, the host halo is not
yet fully assembled and is undergoing a major merger with three other
massive halos.  It is at this epoch, however, that the satellite
experiences the most dramatic tidal mass loss.  Subsequent
tidal peaks result only in a relatively mild stripping.  The object in
the middle panel also suffers a dramatic mass loss at $t\sim 4-5$~Gyr.
In this case, however, the efficient tidal stripping
continues due to the later pericentric passages and associated peaks in
the tidal force.  Finally, the third satellite shown in the figure
experiences only a relatively mild tidal stripping.  This satellite
orbits in the outer regions of the host and never reaches the central
$\approx 60$~kpc.

Note that the pericenter of the third satellite is larger at the late
epochs compared to the pericenter at $t\approx 4$~Gyr.  This is
contrary to a na{\"\i}ve expectation that the pericenter should stay
constant or decrease with time if dynamical friction is efficient.
The real situation is clearly more complicated.  The satellite can
lose as well as gain the orbital energy.  The latter can occur via a
three-body interaction.  Indeed, in examining individual trajectories
we found cases where a satellite gains orbital energy via the
``slingshot'' acceleration --- a classic three-body interaction.

Figure~\ref{fig:tr1} demonstrates that some satellites with small
maximum circular velocity and mass at $z=0$ were substantially more massive
during the early stages of their evolution.  The mass of the object in
the middle column is $\approx 10^{10}\ \rm M_{\odot}$ and its circular
velocity is $>40\ \rm km\,s^{-1}$ at $t=4$~Gyr.  At the present epoch,
they are only $2\times 10^8\ \rm M_{\odot}$ and $18\ \rm km\, s^{-1}$,
respectively.  In the extreme cases we find changes of mass and $V_{\rm
max}$ by a factor of 200 and 8, respectively (see Fig.~\ref{fig:vmt}).

At the same time, the object in the right column of
Figure~\ref{fig:tr1} has a considerably larger pericenter and
experiences weaker tidal force by more than an order of magnitude.
Consequently, its mass and circular velocity change little during the
evolution.  What is the relative frequency of such cases compared to
the cases of dramatic mass loss?  We address this question in the next
section.
%There is therefore a variety of evolutionary histories for 
%a set of surviving satellites in a halo, which depend on the accretion
%time, mass, internal structure and orbital parameters. 

\begin{figure}[t]
\centerline{\epsfysize6.5truein \epsffile{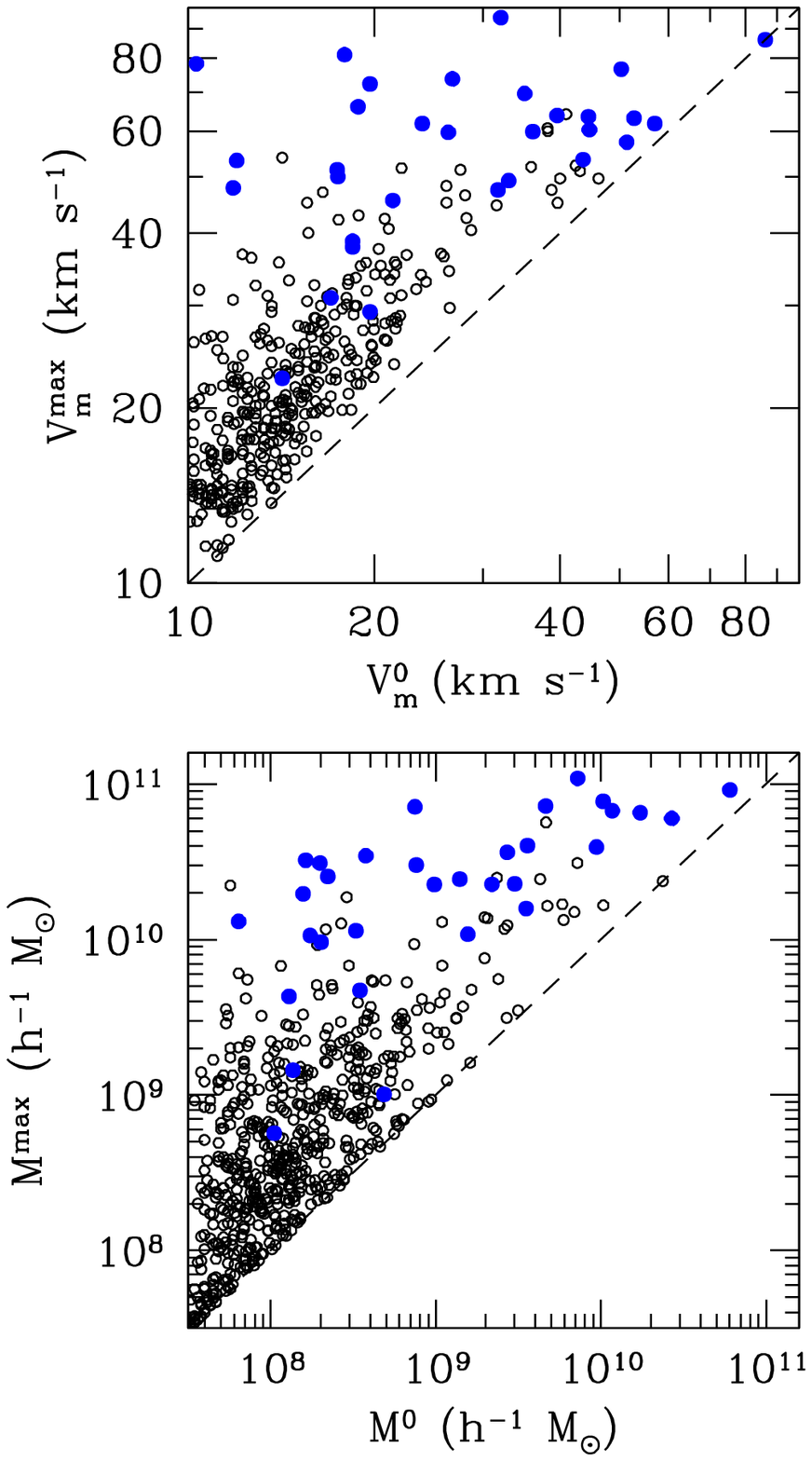}}
\vspace{-0.5cm}
\caption{{\it Top panel:\/} the maximum circular velocity along the trajectory, $V_{\rm m}^{\rm max}$, vs.
  the present-day value $V_{\rm m}^0$. {\it Bottom panel:\/} maximum
  mass along the trajectory vs. the present-day mass of a halo.  The
  satellite halos shown are located within the virial radius of their
  respective host halo. The figure shows that many halos experience a
  dramatic decrease in their mass and circular velocity. The {\it
    solid circles} show the halos that host luminous galaxies in our
  model (see \S~\ref{sec:sam}).
\label{fig:vmmax}}
\end{figure}

%----------------------------------------
\subsection{Internal structure evolution}
%----------------------------------------

Strong tidal forces experienced by orbiting halos lead to a
substantial mass loss, preferentially at the outer radii.  The changes
in the inner regions are more subtle and occur at a slower rate but can
nevertheless be significant
\citep[e.g.,][]{klypin_etal99,hayashi_etal03,stoehr_etal03,kazantzidis_etal03}.
Figure~\ref{fig:vmmax} shows the maximum values of $M$ and $V_{\rm
max}$ reached by a satellite during its evolution vs. their present
values.  Most of the {\it surviving} satellites experience only mild
evolution, less than a factor of two in $V_{\rm m}$.  Yet, there is
a fair number of cases in which the evolution is significant.  The
average changes in $M$ or $V_{\rm m}$ do not seem to depend on the
halo mass.

\begin{figure}[t]
%\vspace{-1cm}
\hspace{-0.3cm}\centerline{\epsfysize4truein \epsffile{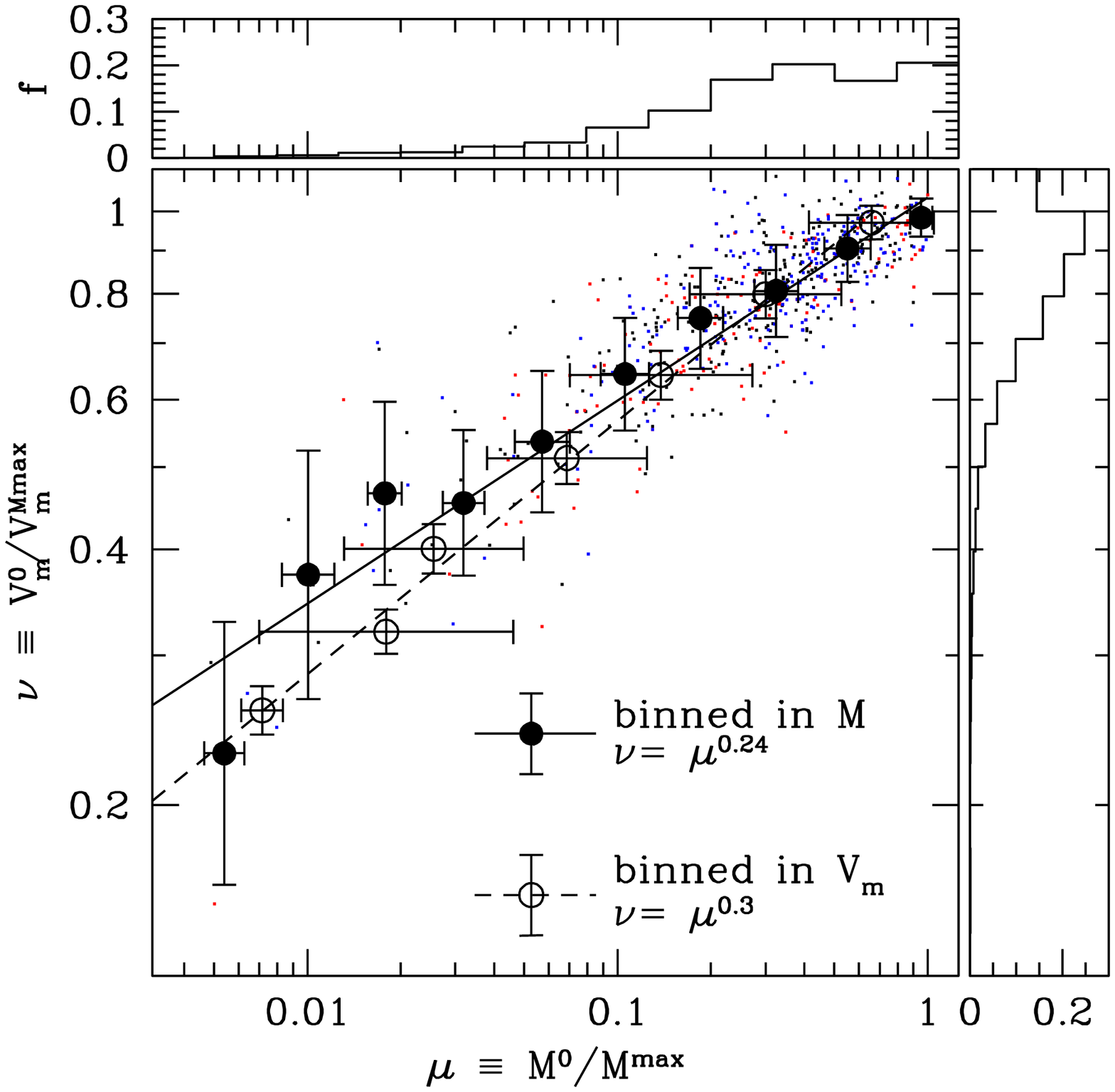}}
\caption{The ratio of the $z=0$ mass 
  to the maximum mass achieved by each satellite during its evolution
  vs. the ratio of the maximum circular velocities at these two
  epochs. The {\it dots} represent the ratios for individual halos.
  {\it Solid circles} show the average for the equally space
  logarithmic mass bins and {\it solid line} shows the power law
  weighted least square fit to these points. {\it Open circles} and
  the {\it dashed line} show the same for the binning in $V_{\rm
    max}$.  The histograms in the {\it top} and {\it right panels}
  show the fraction of the halos with a given mass and circular
  velocity ratio in logarithmic bins of size $0.2$ and $0.05$ for the
  mass and velocity ratios, respectively.
\label{fig:vvmm}}
\end{figure}

\begin{figure}[t]
%\vspace{-1cm}
\centerline{\epsfysize4truein \epsffile{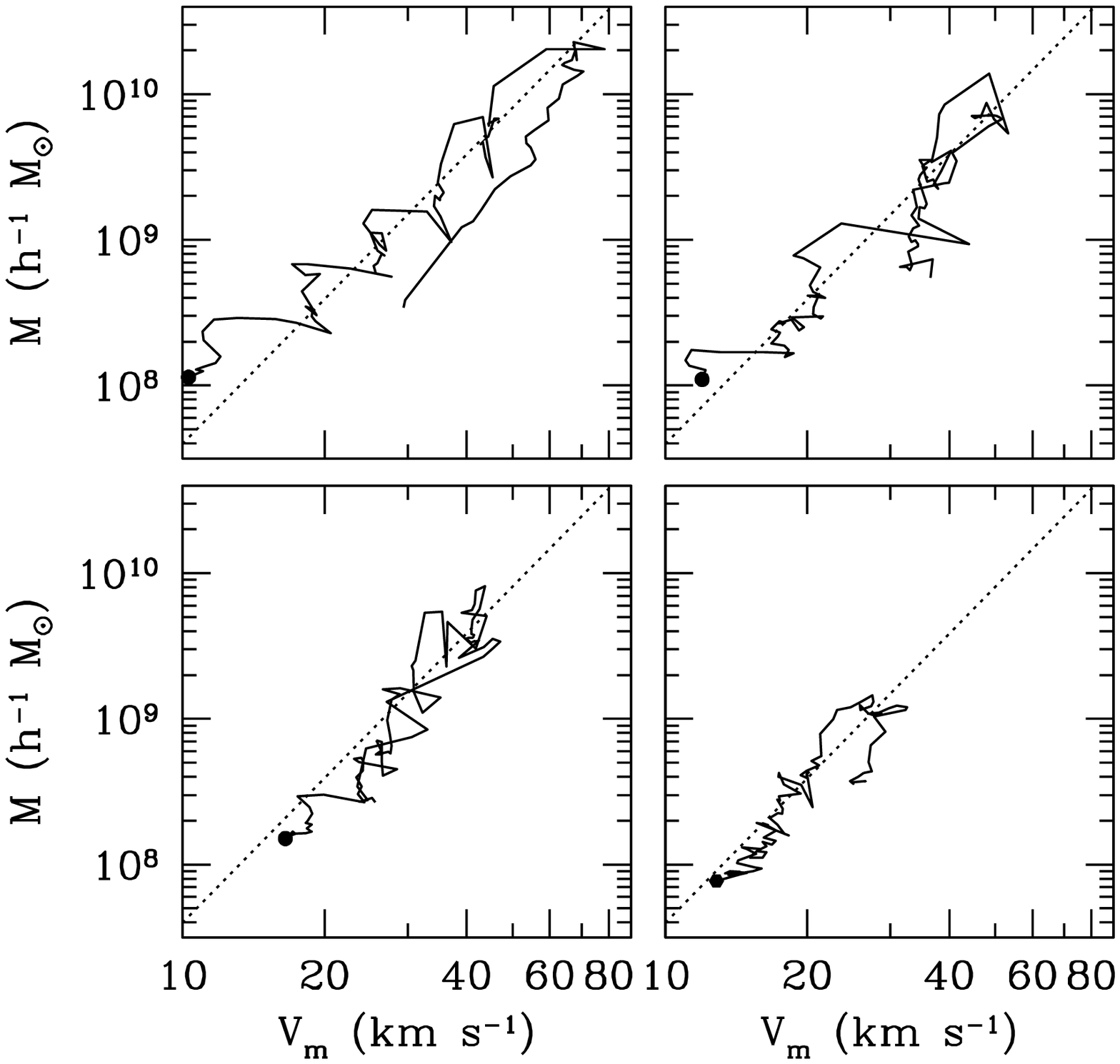}}
\caption{Trajectories ({\it solid lines}) of halos on the  mass --- maximum circular velocity 
  plane for four alos found within the virial radius of all three
  galactic hosts at $z=0$. We selected objects with large changes in
  mass to maximize the dynamic range and illustrate the effect.  The
  solid circles mark the end ($z=0$ epoch) of each trajectory. The
  figure shows that both during the periods of mass growth while
  evolving in isolation and the periods of mass decrease due to tidal
  stripping, halos approximately move up and down the virial
  dependence $M\propto V_{\rm m}^{\alpha}$ with $\alpha\sim 3-5$; the
  {\it dotted line} corresponds to $\alpha=3.3$.
\label{fig:vmt}}
\end{figure}

Figure~\ref{fig:vvmm} shows the ratio of the mass at $z=0$ to the
maximum mass achieved by each satellite during its evolution vs. the
ratio of the maximum circular velocities at these two epochs. The
figure shows a strong correlation between the two ratios:
\begin{equation}
\frac{M^{0}}{M^{\rm max}} = 
  \left(\frac{V_{\rm m}^{0}}{V_{\rm m}^{\rm max}}\right)^{\beta},
  \quad \beta\approx 3-4,
  \label{eq:vvmm}
\end{equation}
where $V_{\rm m}^{0}$ and $M^{0}$ are the values at the
present and $V_{\rm m}^{\rm max}$ is the maximum circular velocity
at the epoch when the halo reached the maximum mass, $M^{\rm
max}$.  This correlation shows that the internal structure of
satellites re-adjusts as they lose mass due to tidal stripping.  Note
that the decrease of $V_{\rm m}$ indicates the decrease in density
within the inner radius of $\approx 2.16r_{\rm s}$, where $r_s$ is the
NFW scale radius.
% of the initial density profile of the satellite.
The adjustment is such that the virial correlation
\begin{equation}
 M\propto V_{\rm m}^{\alpha},\ \ \  \alpha\approx 3-4,
\label{eq:mv}
\end{equation}
is approximately maintained at all times. 

This can be seen in Figure~\ref{fig:vmt}, which shows the tracks of
individual satellites in the $M-V_{\rm m}$ plane.  The
satellites shown were selected from all three galactic hosts.  We
selected objects with large changes in mass to maximize the dynamic
range.  The figure shows that both during the periods of mass growth
while evolving in isolation and the periods of mass decrease due to
tidal stripping, halos approximately move up and down the power-law
dependence of eq.~(\ref{eq:mv}).  For instance, the track shown by
in the top left panel starts at $M\approx 3\times 10^{8}h^{-1}\ \rm M_{\odot}$ and
$V_{\rm m}\approx 30\ \rm km\,s^{-1}$ at $z=10$ and by the redshift
$z=2$ reaches the mass of $2\times 10^{10}h^{-1}\ \rm M_{\odot}$.  In
the ten billion years between $z=2$ and $z=0$, the halo loses $99.5\%$
of the mass and its $V_{\rm m}$ decreases by a factor of eight.
Yet, during the entire course of evolution the halo moves roughly
along the $M\propto V_{\rm m}^{3.3}$ line.

This result is in agreement with \citet{hayashi_etal03}, who found a
correlation similar to that of eq.~(\ref{eq:vvmm}) using controlled
$N$-body experiments to study the tidal stripping and internal
structure of the NFW halos (see their Fig. 12).  They note that the
density at all radii changes in response to tidal shocking.  The
density decrease is greatest at large radii, so that the overall
profile steepens while the normalization drops.  The adjustment of the
density profile leads to the decrease of both $V_{\rm m}$ and $r_{\rm
  max}$.  We also find that $r_{\rm max}$ of satellites in our
simulations decrease systematically as they lose mass.  A similar
evolution of $V_{\rm max}$ as a function of mass loss is found in very
high-resolution controlled simulations of \citet{kazantzidis_etal03},
which followed tidal stripping of an NFW satellite resolved with
$10^7$ particles (Stelios Kazantzidis, private communication).

%----------------------------------------------------------
\subsection{Evolution of halos in the $M - V_{\rm m}$ plane}
%----------------------------------------------------------

In the previous section we showed that individual halos maintain
$M\propto V_{\rm m}^\alpha$ relation during their evolution. This
explains why the same correlation between the mass and $V_{\rm m}$
holds for both subhalos and isolated halos
\citep{avila_reese_etal99,bullock_etal01}.  We also find that the
mass--circular velocity relations for the subhalos at $z=0$ and for
their progenitors at the epoch when the maximum mass was reached have
similar amplitudes and slopes ($\approx 3.3$).

In this section we will consider the mechanism behind such behavior in
more detail.  We can interpret the observed evolution of mass and
$V_{\rm m}$ for a given subhalo by dividing it in the following two
stages: (1) mass growth while evolving in isolation, and
(2) mass decrease due to tidal stripping after the halo is accreted by
its host. The transition between the two stages typically occurs at
$z\sim 2$ for the mass range of subhalos and hosts considered here.

We fit the slope $\alpha$ for the trajectories in the $M-V_{\rm m}$
plane for all satellites separately in the two regimes.  We fit
only tracks of halos with $V_{\rm m} > 15$ km s$^{-1}$ at $z=0$ and
with at least 10 redshift outputs.  In calculating the average slope,
$\alpha_0$, and the dispersion of the sample, $\sigma_\alpha$, we
weigh $\alpha$ of each halo by the error of the fit.  The isolated
halos have the average slope $\alpha_0 = 4.7$, with the dispersion
$\sigma_\alpha = 1.2$.  The truncated halos have $\alpha_0 = 2.9$ and
$\sigma_\alpha = 1.2$.  Thus the slopes in the two regimes seem to be
somewhat different.

These different slopes can be linked to the different average
densities of the dwarf halos in the two regimes.  In the mass-growth
stage, when the average density of the Universe is
$\bar{\rho}(z)=\rho_0(1+z)^3$, the mass and velocity are given by the
virial scaling relation $M_{\rm vir} \propto V_{\rm vir}^3
(1+z)^{-3/2}$.  Also, \citet{bullock_etal01} showed that, as long as
the NFW model is an adequate description of the halo profile, $V_{\rm
m}$ and $V_{\rm vir}$ are related through the concentration parameter
approximately as $V_{\rm m}/V_{\rm vir} \propto c_{\rm vir}^{1/4}$.
The median concentration itself varies with the mass and redshift as
$c_{\rm vir} \propto (1+z)^{-1} M_{\rm vir}^{-0.13}$.  Since all the
variables scale as some power of $(1+z)$, it is natural to approximate
the evolution of the mass and maximum velocity as $M_{\rm vir} \propto
(1+z)^{-q}$ and $V_{\rm m} \propto (1+z)^{-p}$, with the above
relations leading to $\frac{29}{32}q = 3p + \frac{3}{4}$.  The slope
$\alpha_0 = q/p = 4.7$ is achieved for $q=2.8$ and $p=0.6$, although
the scatter in the value of $\alpha$ implies a corresponding scatter
in the exponents $q$ and $p$.

Note that the slope of the $M_{\rm vir} - V_{\rm m}$ relation is
steeper than the virial $\alpha \approx 3$ because the virial
parameters of isolated halos depend on the mean density of the
Universe and that density is changing with redshift.  The same
zero-point of the relation can be maintained only if both $M_{\rm
vir}$ and $V_{\rm m}$ are changing with time in a certain way,
specified above.

During the second stage of evolution, the subhalo experiences tidal
forces from the host and other halos, and its mass and extent are
tidally truncated. The average density $\rho_{\rm t}$ within the
truncation radius $R_{\rm t}$ is approximately constant along the
orbit and is proportional to the background density of the host halo
at the pericenter of subhalo's orbit.  The truncation radius scales
with the truncated mass, $M_{\rm t}$, as $R_{\rm t} \propto (M_{\rm
t}/\rho_{\rm t})^{1/3}$, so that $M_{\rm t} \propto V_{\rm t}^3
\rho_{\rm t}^{-1/2}$.  The velocity $V_{\rm t} \propto (M_{\rm
t}/R_{\rm t})^{1/2}$ is a good estimate for the peak velocity when the
subhalo is severely truncated.  If the background density is constant
along the satellite trajectory, we obtain the following relation:
$M_{\rm t} \propto V_{\rm m}^3$, in agreement with the average fit
in this regime.  Figure \ref{fig:vmt} shows that in this regime the
power-law relation between $M_t$ and $V_{\rm m}$ has a significant
dispersion, which is due to the variation of $\rho_t$ along the
trajectory.  Nevertheless, as long as the distance of closest approach to the host
halo remains the same, the average relation is well
maintained.

Thus, we expect the mass-velocity relation to be constrained by the
two limiting slopes 3 and 5.  The actual mass accretion and mass loss
history may vary from halo to halo but the same $M - V_{\rm m}$
relation is maintained throughout the evolution, with a transition
from the initial slope $\alpha=4.7$ for the isolated halos to the
later slope $\alpha=3$ for the tidally truncated halos.

%---------------------------------------------------------------------
\subsection{Abundance and radial distribution of galactic satellites}
  \label{sec:vfrd}
%--------------------------------------------------------------------

\begin{figure}[t]
%\vspace{-1cm}
%\centerline{\epsfysize4truein \epsffile{vflg_400.eps}}
\centerline{\epsfysize4truein \epsffile{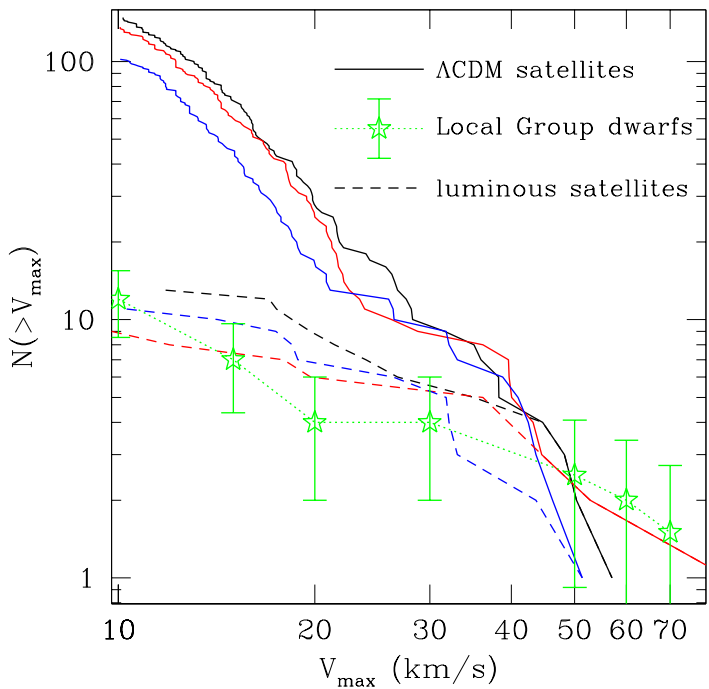}}
\caption{The cumulative velocity function of the dark matter 
  satellites in the three galactic halos ({\it solid lines\/} compared
  to the average cumulative velocity function of dwarf galaxies around
  the Milky Way and Andromeda galaxies ({\it stars}). For the objects
  in simulations $V_{\rm circ}$ is the maximum circular velocity,
  while for the Local Group galaxies it is either the circular
  velocity measured from rotation curve or from the line-of-sight
  velocity dispersion assuming isotropic velocities. Both observed and
  simulated objects are selected within the radius of $200h^{-1}\ \rm
  kpc$ from the center of their host. The dashed lines show the
  velocity function for the luminous satellites in our model described
  in \S~\ref{sec:sam}.  The minimum stellar mass of the luminous
  satellites for the three hosts ranges from $\approx 10^5\ \rm
  M_{\odot}$ to $\approx 10^6\ \rm M_{\odot}$, comparable to the
  observed range.
\label{fig:vflg}}
\end{figure}

  Figure~\ref{fig:vflg} shows the cumulative velocity functions
  (CVFs), the number of satellites with maximum circular
  velocity\footnote{Note that uncertainty in the velocity anisotropy
    affects the conversion of the line-of-sight {\it rms\/} velocity
    of dSph galaxies to $V_{\rm m}$.  In the plot we assume an
    isotropic velocity distribution.  Our re-analysis of numerical
    simulations of \citet{gnedin03b} shows that tidal truncation and
    heating of galaxies leads to the preferential removal of radial
    orbits and the development of the tangentially-biased dispersion
    in the outer parts.  A similar result has been found by
    \citet{kazantzidis_etal04a} and \citet{moore_etal03}.  The
    solution of the Jeans equation for $V_{\rm m}$ is sensitive to the
    exact value of the anisotropy parameter
    \citep{zentner_bullock03,kazantzidis_etal03}.} larger than a given
  value, for the objects located within $200h^{-1}\ \rm kpc$ of their
  host halo.  The figure compares the CVFs for the DM satellites and
  observed satellites of the MW and Andromeda
\footnote{We use the circular velocities compiled by
\citet{klypin_etal99a} with updated values of circular velocity for
the Large and Small Magellanic Clouds of $V_{\rm m}=50\rm\
km\,s^{-1}$ and $60\rm\ km\,s^{-1}$, respectively
\citep{vandermarel_etal02}}
and highlights the ``missing satellite problem''
\citep{kauffman_etal93,klypin_etal99a,moore_etal99}: a large difference in the number
of dwarf-size DM satellites in simulations and the observed number of
dwarfs in the Local Group.

\begin{figure}[t]
%\vspace{-1cm}
\centerline{\epsfysize4truein \epsffile{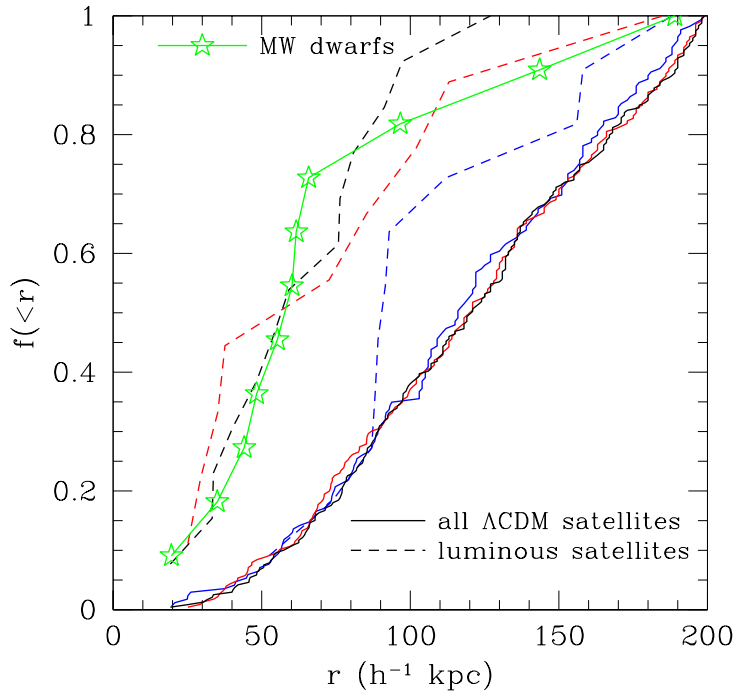}}
\caption{The fraction of satellites within a certain distance from the
  center of their host galaxy. The {\it solid lines\/} show distributions of
  the $\Lambda$CDM satellites in the three galactic halos, while the
  {\it connected stars} show the distribution of dwarf galaxies around the
  Milky Way. The figure shows that radial distribution of observed
satellites is more compact than that of the overall population of dark 
matter satellites. The {\em dashed} lines show distributions for the 
luminous satellites in our model (\S~\ref{sec:sam}). The population
of luminous satellites is the same in this and previous figures. 
\label{fig:rdlg}}
\end{figure}

Figure~\ref{fig:rdlg} shows the normalized cumulative radial
distribution of the DM satellites compared to the radial distribution
of satellites around the Milky Way within the same radius. The Local
Group data is from the compilation of \citet{grebel_etal03}.  The
figure clearly shows that the spatial distribution of dwarf galaxies
around the Milky Way is more compact than the distribution of the DM
population.  The median distance of observed satellites within
$200h^{-1}$~kpc is $60h^{-1}$~kpc and $85h^{-1}$~kpc for the MW and
M31, respectively.  For the DM satellites the corresponding median
distances are $116h^{-1}$ kpc, $121h^{-1}$ kpc, and $120h^{-1}$
kpc.  Although the median for M31 satellites is smaller than that of
the DM satellites, their radial distributions are formally
consistent.  However, the comparison with the M31 satellites is
difficult at present because typical distance errors are $\sim
20-50$~kpc (and $\gtrsim 70$~kpc for some galaxies), comparable to the
distance to the host.

For the MW satellites the typical distance errors are an order of
magnitude smaller and the comparison is considerably more
meaningful. The Kolmogorov-Smirnov (KS) test gives probability of
$(6-8)\times 10^{-4}$ that the MW satellites are drawn from the same
radial distribution as the DM satellites.  This has also been pointed
out recently by \citet{taylor_etal03}, who compared the spatial
distribution of the MW satellites to results of their semi-analytic
model of galaxy formation.  Thus, in addition to the vastly different
abundances of the observed and predicted satellites, there is a
discrepancy in the radial distribution.  Models that aim to reproduce
the abundance of the LG satellites should therefore be able to
reproduce the radial distribution as well.

%------------------------------------------------------------------
\section{A model of star formation in dwarf halos}
  \label{sec:sam}
%------------------------------------------------------------------

\subsection{Description of the Model}
\label{sec:sfmodel}

In order to gain insight into which halos might become luminous and
which might not, we implement the following simple model of star
formation.  We use the standard assumption that the gas within the
halos with the virial temperature $T_{\rm vir} > 10^4$ K dissipates
its energy via radiative cooling and forms a disk.  We then apply the
empirical Schmidt law to calculate the star formation rate in radial
shells within the disk.  The novel features of our model include: (i)
use of mass accretion and stripping history of the dwarf halos
extracted from simulation, (ii) effects of photoionizing extragalactic
background using the filtering mass, (iii) effects of inefficient
dissipation of the gas at $T_{\rm vir} \lesssim 10^4$ K, and (iv)
bursts of star formation due to strong tidal shocks. The details of 
the model are as follows.

(i) Using the mass assembly history (MAH) of a given satellite halo directly
from the simulation, instead of a semi-analytic approach, we are able
to trace the major merger events as well as the quiescent accretion of
material.  The halo mass increases in both regimes, but the star formation
rates are very different. The use of simulation MAHs allows us to determine
the accretion epoch of a satellite and follow its mass loss due
to tidal stripping.

(ii) At each time output we calculate the accreted mass since the last
time step, $\Delta M$.  We increase the total gas mass, $M_g$, in the
satellite by the amount of cold gas in a single halo with the mass
$\Delta M$ at that epoch: $\Delta M_g = f_g(M,z) \Delta M$.  The
fraction $f_g$ takes into account the photoevaporation of baryons by
extragalactic UV flux, using the filtering scale parametrization of
\citet{gnedin00} and taking the redshift of reionization to be $z_r =
7$.  See Appendix \ref{sec:filter} and equation (\ref{eq:fg}) for
details.  After the satellite enters the host halo, the accretion of
new gas is halted and the disk scale length is fixed, although stars
may continue to form from the remaining reservoir of cold gas.

We distribute the gas on a spherically symmetric grid of 50 radial
shells, according to the surface density of an exponential disk:
$\Sigma_g(r) = \Sigma_0 \exp{(-r/r_{\rm d})}$.  We use the observed Schmidt
law of star formation to estimate the star formation rate:
$\dot{\Sigma}_* = 2.5\times 10^{-4} \, (\Sigma_g / M_{\sun} \,
\mbox{pc}^{-2})^{1.4} \, M_{\sun} \, \mbox{kpc}^{-2}$.  Only the
shells above the threshold $\Sigma_g > \Sigma_{\rm th} \equiv 5 \
M_{\sun} \, \mbox{pc}^{-2}$ form stars \citep{kennicutt98}.

(iii) The scale length of the disk is determined by its angular
momentum.  For a rotationally-supported disk it is approximately $r_{\rm d}
= 2^{-1/2} \lambda \, r_{\rm vir}$.  The value of the angular momentum
parameter is drawn randomly from the probability distribution
\begin{equation}
  p(\lambda) d\lambda = {1 \over \sqrt{2\pi} \sigma_\lambda}
     e^{-{(\ln{\lambda/\bar{\lambda}})^2 \over 2 \sigma_\lambda^2}}
     {d\lambda \over \lambda},
  \label{eq:lambda}
\end{equation}
with $\bar{\lambda} = 0.045$, $\sigma_\lambda = 0.56$, according to
the latest measurement by \cite{vitvitska_etal02}.  This is a key
assumption of the semi-analytic models of galaxy formation.

However, small halos at high redshift could only cool by atomic
hydrogen to about $10^4$ K.  If their virial temperature is only
slightly above that equilibrium temperature, the gas would not be able
to dissipate enough to reach a rotationally-supported state.  Instead,
its distribution would be more extended, which can have important
implication for the star formation with a density threshold
$\Sigma_{\rm th}$.  This effect is particularly important for dwarf
halos.

We model the effect of inefficient dissipation by adopting the
expansion factor that depends on the ratio of the virial temperature
to the equilibrium temperature of $10^4$ K.  The gas would reach a
Boltzmann distribution with the density $M/r^3 \propto
\exp{(-\Phi/kT)}$, where $\Phi$ is the potential energy.  Using the
maximum circular velocity instead of the temperature and ignoring the
slow variation of the potential, we can express the scale length of
the gas as $r_{\rm d} \propto \exp{\left[c (V_4/V_{\rm m})^2\right]}$,
where $c$ is a normalization factor and $V_4 = 16.7$ km s$^{-1}$ is
the virial velocity corresponding to $T_{\rm vir} = 10^4$ K.  We find
that $c = 10$ is a best fit to the abundance and radial distribution
of the Local Group galaxies (see \S\ref{sec:sfresults}). This scaling
also provides a good description of the extent of the gas within halos
in cosmological galaxy formation simulation described in
\citet{kravtsov_gnedin03}.  Thus, we set the size of the gaseous disk
at each time step to be
\begin{equation}
  r_{\rm d} = 2^{-1/2} \lambda \, r_{\rm vir} \times e^{10 (V_4/V_{\rm m})^2}.
\label{eq:rd}
\end{equation}
Of course, $r_{\rm d}$ is not allowed to exceed the tidal radius of the
halo, $r_{\rm t}$.  The gas in large halos with $V_{\rm m} \gg V_4$ can
cool efficiently and reach rotational support, but for small halos
with $V_{\rm m} \gtrsim V_4$ the extended distribution reduces the
central concentration of the gas and hinders star formation.

(iv) Strong tidal forces, such as in the interacting or merging
galaxies, may lead to a burst of star formation throughout the dwarf
galaxy.  The association of starbursts with strong peaks of the tidal
force is motivated by theoretical models \citep{mayer_etal01b} and, to
a certain extent, by observations \citep{zaritsky_harris03}.  The
latter suggest that the tidally-triggered star formation in the SMC
can be accurately modeled as an instantaneous burst of star formation.
\citet{zaritsky_harris03} find the best fit to their data when the
star formation rate (SFR) varies as $r^{-4.6}$ with the distance to
the Galaxy.  The tidal interaction parameter, $I_{\rm tid}$ (see
eq. [\ref{eq:Itidsum}]), that reflects the integrated effect of a
single tidal shock, is the most natural candidate for the
parametrization of the tidally-triggered SFR.  Ignoring the adiabatic
correction, it varies with the distance to the perturber approximately
as $I_{\rm tid} \propto r^{-4}$ (but see the discussion in
\S~\ref{sec:sfresults}).

We allow for the starburst mode of star formation, when the tidal
interaction parameter exceeds a threshold value.  After experimenting with
different thresholds, we find that $I_{\rm tid,th} = 4\times 10^3$
Gyr$^{-2}$ provides the best simultaneous fit to the velocity function
and spatial distribution of the satellites.  In all radial shells, a
fraction $f_* = {I_{\rm tid} / 4\times 10^4 \ \mbox{Gyr}^{-2}}$ (with
a maximum of $f_* = 0.5$) of the available gas is converted into stars
instantaneously.  The normalization of $f_*$ is somewhat arbitrary and
can be adjusted to fit the stellar masses of the satellites.  Since
the starburst changes drastically the distribution of gas in the
galaxy, new infalling gas may have a very different angular momentum.
Therefore, after each starburst we recalculate the value of $\lambda$
according to eq. (\ref{eq:lambda}).

The external tidal force determines the truncation radius $R_{\rm t}$ of the
satellite, outside which all stars and gas are lost.  In a static
gravitational field, the radius of the Roche lobe is set by the
condition that the average density of matter in the satellite equals
twice the local ambient density (for the isothermal sphere potential).
In a dynamic situation of the satellite on an eccentric orbit
experiencing tidal shocks, the truncation depends on the
time-varying tidal force.  However, using $N$-body simulations of the
dynamical evolution of galaxies in clusters, \citet{gnedin03b} showed
that the truncation radius can be accurately described by the same
condition, $\rho_{\rm av}(R_{\rm t}) = 2 \, \rho_{\rm tid}$, where the
effective tidal density $\rho_{\rm tid}$ is related to the trace of
the tidal tensor via
\begin{equation}
\rho_{\rm tid} = 1.8 \times 10^{-5} \, \left({F_{\rm tid} \over
   \mbox{Gyr}^{-2}}\right) \; M_{\sun}\, \mbox{pc}^{-3}.
\end{equation}
The truncation occurs near the maximum of the tidal force along the
orbit, usually at the perigalactic distance.

The knowledge of the external tidal force also allows us to estimate
the tidal heating of stars in the satellite.  After each tidal shock,
typically once per orbit, the velocity
dispersion of stars in each radial shell increases by the amount
\citep{gnedin03a}
\begin{equation}
  \Delta \sigma^2(r) = 0.32 \left({I_{\rm tid} \over \mbox{Gyr}^{-2}}\right)
     \left({r \over \mbox{kpc}}\right)^2 
     \ \mbox{km}^2 \, \mbox{s}^{-2}.
  \label{eq:sigma}
\end{equation}
The mass-weighted dispersion $\sigma$ may serve as an indicator of the
morphological type of the satellite.  In \S \ref{sec:sfresults}, we
adopt the ratio of the rotation velocity to the velocity dispersion,
$v_{\rm rot}/\sigma$, as a possible criterion.  In practice, we
compute $v_{\rm rot}$ as the circular velocity of the NFW halo at the
radius enclosing all bound stars.

%--------------------
\subsection{Results}
\label{sec:sfresults}
%--------------------

\begin{figure}[t]
%\vspace{-1cm}
\centerline{\epsfysize3.75truein \epsffile{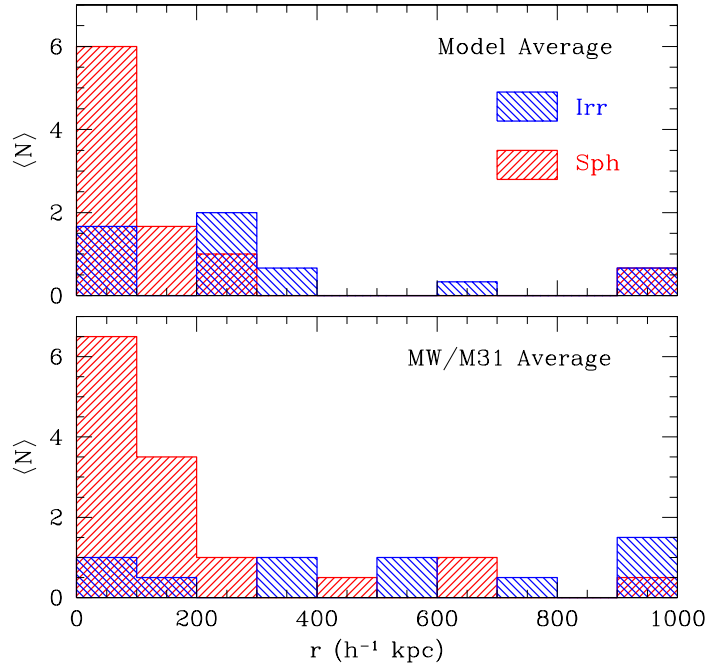}}
\caption{Morphological segregation of galaxies in the Local Group
  ({\em lower panel}) and in our model ({\em upper panel}). We divided
  the observed galaxies in two broad classes: Irr -- all dwarf
  irregular galaxies, and Sph -- all spheroidal systems, including
  dSph, dSph/dIrr, and dEs. The model result include all subhalos that
formed stars and the division into irregular and spheroidal systems
was done using the amount of heating experienced by each object, as 
explained in \S~\ref{sec:sfresults}. The model reproduces the observed 
morphological segregation: most spheroidal systems are located within
$300h^{-1}$~kpc of the host, while irregular systems are found at 
a wide range of radii. 
\label{fig:mor}}
\end{figure}

We show the predictions of our model for the cumulative velocity
function and radial distribution of luminous satellites by dashed
lines in Figures~\ref{fig:vflg} and \ref{fig:rdlg}.  The model
reproduces fairly well both observational statistics. The abundance of
luminous satellites and the shape of the velocity function are in
reasonable agreement with observations.  The stellar masses of the
satellites are in the range $10^5\lesssim M_{\ast} \lesssim 10^{10}\ 
\rm M_{\odot}$, similar to the observed range of stellar masses of the
LG galaxies \citep[e.g.,][]{dekel_woo03}.  As can be seen in
Figure~\ref{fig:vmmax}, all of the luminous systems in our model,
including those that have masses and circular velocities of the
smallest dwarfs, were relatively massive ($M\gtrsim 10^9\rm\ 
M_{\odot}$ and $V_{\rm m}\gtrsim 30\rm\ km\,s^{-1}$) at some point in
their evolution.

In comparison to the observed velocity functions it is worth noting
that the conversion between the line-of-sight stellar velocity dispersions
and maximum circular velocity is somewhat uncertain
\citep{stoehr_etal02,zentner_bullock03,kazantzidis_etal03}.  Thus, at
this point it is not worth trying to reproduce the observed function
exactly. 

The median distances of luminous satellites to their respective hosts
are $59h^{-1}$~kpc, $91h^{-1}$~kpc, and $73h^{-1}$~kpc for halos
G$_1$, G$_2$, and G$_3$, respectively.  This is close to the median
values for the MW and M31 and lower than the median distance of the
overall DM satellites ($\approx 120h^{-1}$ kpc, see
\S~\ref{sec:vfrd}). The KS probability that the radial distribution of
these luminous satellites is drawn from the same distribution as
that of the MW are 97\%, 1\%, and 75\% for the three hosts,
respectively. Although there are apparent fluctuations due to the
differences in the evolutionary histories of the three hosts, the
luminous satellites in our model have a clear tendency to be more
centrally concentrated than the overall DM satellite population.

On the other hand, tidally-triggered bursts of star formation are not
limited to the central parts of the host halo.  In the isolated halo
G1, where the sample of satellites is not contaminated by the close
proximity to another host, the tidal heating parameter $I_{\rm tid}$
scales with the present distance as $I_{\rm tid} \propto r^a$, $a =
-3.7 \pm 0.2$, for $r < 1 \, h^{-1}$ Mpc, where the quoted error is
the standard deviation of the whole sample, not the error of the mean.
This is consistent with the expected slope $a = -4$
(c.f. eq. [\ref{eq:Itid}], ignoring the adiabatic correction).
However, if we limit the sample to the halos of interest, i.e. only
those capable of forming stars (with the maximum $V_{\rm m} > V_4 =
16.7$ km s$^{-1}$, the virial velocity corresponding to $T_{\rm vir} =
10^4$ K), then the slope is significantly shallower: $a = -1.8 \pm 0.2$.
Furthermore, if we consider only the largest tidal parameters that
might lead to starbursts ($I_{\rm tid} > 10^3$ Gyr$^{-2}$), then the
distribution is almost independent of distance: $a = -0.3 \pm 0.1$.
Thus, the current location of the satellite in the host galaxy gives
very little indication whether it had tidal starbursts in the past.

We find a large variety of star formation histories for the luminous
satellites.  Most systems have a single initial burst lasting up to 2
Gyrs. For some this is the only starforming activity, while other have
a constant SFR at $1 \, \rm M_\odot$ yr$^{-1}$ up to $z=0.7$ or bursty
star formation continuing until $z=0.2$. There are also objects that
have only a single tidally-triggered burst at $z\sim 1$. The tendency
is for more massive satellites to have more extended star
formation.  The median mass-weighted epoch of star formation in halos
with $V_{\rm m} < 30$ km s$^{-1}$ is between $t_{\rm med} = 1-4$ Gyr
cosmic time (corresponding redshifts $z = 5 - 1.7$), while the more
massive halos have $t_{\rm med}$ up to 7 Gyr ($z = 0.7$).

Our simple model also predicts central stellar densities in a
reasonable agreement with observations: roughly constant
$\Sigma_{\ast} \sim 5-50 \rm\ M_{\odot}\, pc^{-2}$ for $M_{\ast} <
10^9 \rm\ M_{\odot}$ and rising with the stellar mass as
$\Sigma_{\ast} \sim M_{\ast}/(10^7\rm\ M_{\odot}) \rm\ M_{\odot} \,
pc^{-2}$ for systems with $M_{\ast} > 10^9\rm\ M_{\odot}$.  The
satellites located within 100 kpc of their host galaxy have typically
higher central densities ($\gtrsim 50 \rm\ M_{\odot}\, pc^{-2}$) than
the more distant satellites.

The results listed above are valid for all dwarf satellite galaxies
regardless of their evolutionary history.  In addition, as discussed in the
previous section, our model tracks the tidal heating of stars formed
in each halo.  We can therefore attempt a crude morphological
classification of galaxies based on the amount of heating they
experienced.  This is motivated by the observations that dSph galaxies
have low values of the ratio of rotation velocity to the random
velocity dispersion, $v_{\rm rot}/\sigma\lesssim 1$.  The galaxies of
transition type dIrr/dSph have $v_{\rm rot}/\sigma\lesssim 2$
\citep[see, e.g.,][and references therein]{grebel_etal03}.
Theoretical models of \citet{mayer_etal01a,mayer_etal01b} also
indicate that the tidally-heated dSph-like remnants of low-surface
brightness spiral galaxies have small $v_{\rm rot}/\sigma$.

We use the circular velocity at the radius enclosing all of the
stellar mass, $v_{\rm c}^{\rm out}$, as a proxy for $v_{\rm rot}$. The
rotation velocity will, in general, be smaller than the circular
velocity because some of the kinetic energy is in the form of the
random motions. Also, we account only for direct tidal heating and do
not take into account tidally-induced heating via bar and bending
instabilities. The exact value of $v_{\rm rot}/\sigma$ for our
galaxies is thus somewhat uncertain as our $\sigma$ may be regarded as
lower limit.  We experimented with several values for the
classification threshold in the range $1 < v_{\rm c}^{\rm out}/\sigma
< 3$, but the main trends are not sensitive to the specific choice in
this interval. We chose the value of $v_{\rm c}^{\rm out}/\sigma = 3$
for the classification shown in Figure~\ref{fig:mor}. For the observed
galaxies, we combined dwarf spheroidal, transition type, and dwarf
elliptical galaxies in one broad class of spheroidal systems, using
Table~1 of \citet{grebel_etal03}.  The figure shows that our model is
consistent with the observed trend of morphological segregation. Most
spheroidal systems are located within $300h^{-1}$~kpc of the host,
while irregular systems are found at a wide range of radii. The two
model spheroidal systems at $\sim 1h^{-1}$~Mpc have been part of a
small group of galaxies and were tidally heated within this group
before it was accreted by the host.

%---------------------
\section{Discussion}
\label{sec:discussion}
%--------------------

In the previous sections we presented the results of the dynamical
evolution of galactic satellites in self-consistent cosmological
simulations.  One of the main findings is that the internal structure
of the satellites responds to the changes of mass in a remarkably
regular way.  Namely, both during the periods of mass growth and tidal
mass loss, the maximum circular velocity of a halo changes as $V_{\rm
  m}\propto M^{1/\alpha}$. The slope $\alpha$ is $\sim 4-5$ when the
mass grows and $\alpha\approx 3$ when the mass decreases due to tidal
stripping.  The latter result was also obtained by
\citet{hayashi_etal03} in their non-cosmological simulations of
satellite evolution.

The overall evolution of subhalo population is such that their
$M-V_{\rm m}$ relation is similar to that of isolated halos (with
$\alpha\approx 3.3$).  The fact that isolated halos and subhalos have
similar mass-circular velocity relations may hint at why the
fundamental plane of galaxies in clusters and the field are similar
\citep{dressler_etal87,djorgovski_davis87, mobasher_etal99,
bernardi_etal03} and why the scatter in the Tully-Fisher relation is
so small \citep{kannappan_etal02}.

The fact that the circular velocity decreases with decreasing mass
means that the systems experiencing dramatic mass loss will experience
a significant change in circular velocity. We find that about 10\% of the subhalos with masses $<10^8-10^9\ \rm
M_{\odot}$ or $V_{\rm m}<30\ \rm km\,s^{-1}$ at $z=0$ have
considerably larger masses and circular velocities at earlier epochs.
This may explain how such apparently small objects like Ursa Minor and
Draco could have formed stars, given that the gas accretion is expected to
be strongly suppressed by the UV background
\citep[e.g.,][]{thoul_weinberg96,gnedin00}. In our model, these
systems were once sufficiently massive ($V_{\rm m}\gtrsim 30\ \rm
km\,s^{-1}$) to accrete gas and form stars but the accretion was
halted when they started to experience tidal mass loss.

After the accretion of new gas stops, these systems may continue to
form stars in bursts as they are tidally stirred
\citep[e.g.,][]{mayer_etal01a}. Interestingly, we find that the strongest
tidal interaction may occur even before halo is accreted by the host.  Some
satellites experience the strongest tidal force from multiple halos at
early epochs in major mergers during the assembly of their host 
(see Fig.~\ref{fig:tr1}). Such mergers are frequent at early epochs, 
and we find that in general all satellites forming
stars experience multiple bursts in the first $2-3$ Gyrs of their evolution.
We present a simple model for star formation in dwarf halos and apply
it to the evolutionary tracks extracted from the simulations.  As
shown in Figure~\ref{fig:vflg}, the model is successful in reproducing
the abundance of luminous satellites around M31 and the Milky Way.

The spatial distribution of dwarfs around the Milky Way offers another
independent challenge to any model of satellite
evolution. Figure~\ref{fig:rdlg} shows that our model reproduces the
observed distribution reasonably well. The distribution of luminous
satellites is more compact than the overall population of subhalos
because stars form only in objects that were sufficiently massive at
high redshifts.  Due to the strong mass- and redshift-dependence of spatial
bias, such objects are considerably more clustered around the host
than smaller halos that form at a wide range of
redshifts. Correspondingly, we find that luminous objects were
accreted by the host systematically earlier (by $\Delta z\sim 0.5-1$)
than smaller mass dark subhalos.

One of the remarkable features of our model is that the results are
not sensitive to the details of reionization history of the Universe.
For example, all of the presented results are nearly intact if we
change the assumed redshift of reionization from the fiducial value of
$z_{\rm r}=7$ to $z_{\rm r}=15$ (see \S~\ref{sec:filter}). The
physical reason behind this insensitivity to reionization is the
inefficiency of gas cooling and star formation in small mass ($T_{\rm
  vir}\lesssim 10^4$~K) systems. This is because gas in such systems
cannot cool via hydrogen line emission and must rely on the
inefficient H$_2$ cooling. Such redshift-independent suppression of gas
cooling is observed in cosmological simulations of \citet{chiu_etal01}
and \citet{kravtsov_gnedin03}. The important implication is that
properties of the population of galactic satellites are determined by
the physics of galaxy formation rather than by the UV background and
reionization.

Our results can qualitatively explain the morphological segregation of
the Local Group galaxies \citep[e.g.,][]{grebel00}. As shown in
Figure~\ref{fig:mor}, a simple division of model galaxies into
irregular and spheroidal based on the amount of tidal heating they
experienced during their evolution reproduces the main observed trend.
Most spheroidal systems are located within $300h^{-1}$~kpc, while
irregular galaxies are found almost uniformly at all distances.
Therefore, our results support the scenario that spheroidal systems
form via strong tidal heating \citep{mayer_etal01a,mayer_etal01b}.
Note, however, that tidal heating is not restricted to the host. It
can occur early on, before the host is assembled, within merging
subgroups.

Interestingly, this explains a puzzling presence of the Cetus and
Tucana dSph galaxies at the outskirts of the Local Group some
$700$~kpc and $1000$~kpc from the nearest massive spiral (MW or M31).
We also find $\sim 1-2$ galaxies with significant heating (the ratio
of the rotational velocity to the velocity dispersion of $v_{\rm
rot}/\sigma < 1$) at distances $\sim 1000$~kpc from their hosts.  The
tidal heating of these systems occurred in small groups that are being
accreted by the host at the current epoch \citep[see also][for a
similar effect in clusters of galaxies]{gnedin03a}.  As the tidal
force unbinds satellites from such accreting groups, isolated dSph
galaxies may be found at large distances from the primary host.

Also, early tidal interaction, experienced for example by the system shown
in the left column of Fig.~\ref{fig:tr1}, and subsequent interaction
with other subhalos may lead to the increase of orbital energy and
apocenter distance.  This scenario would also explain presence of
dSphs at large distances from the primary.  The main point in both
scenarios is that primary is not the only source of tides and the
present-day environment is not necessarily indicative of a dwarf
galaxy's past.

One of the most interesting candidates for the ``missing'' dark halos
is the population of compact high-velocity clouds (CHVCs) of neutral
hydrogen \citep[HI, e.g.,][]{blitz_etal99,braun_burton99}.  This idea
has recently been boosted by the detection of concentration of CHVCs
near M31 \citep{thilker_etal03}.  It is thus interesting to consider
the amount of gas associated with the subhalos that remain dark in our
model.  The cumulative gas mass function associated with dark halos is
remarkably consistent for all three host halos: $N(>M_{\rm g})\approx
20(M_{\rm g}/10^7 \, \rm M_{\odot})^{-0.7}$ for $10^6<M_{\rm
g}\lesssim 10^8\rm\ M_{\odot}$ within $200h^{-1}\rm\ kpc$.  Most of
the gas mass is thus in most massive subhalos.  The total mass of gas
associated with such halos within $200h^{-1}\rm\ kpc$ is $M_{\rm
g}^{\rm tot}\approx 2\times 10^9\rm\ M_{\odot}$, the number similar
for all three hosts.  If we assume that on the average about $10\%$ of
gas is neutral \citep{maloney_putman03,thilker_etal03}, the total mass
in neutral hydrogen is $M_{\rm HI}\approx 2\times 10^8\rm\ M_{\odot}$.

The predicted number of dark clouds with $M_{\rm g}>10^6 \, \rm
M_{\odot}$ is $\sim 50-100$.  A fraction of the observed CHVCs can
thus be associated with the small-mass DM halos. Within central
$50$~kpc, however, the number of halos with such gas masses is only
$\sim 2-5$.  We cannot therefore explain 25 CHVCs observed by
\citet{thilker_etal03} within this radius around M31.  It is possible
that simulations underpredict the number of small-mass halos due to
overmerging. To check this will require higher-resolution simulations.
On the other hand, we did not take into account processes such as ram
pressure stripping, which would further reduce the number of halos
with gas.  Another possibility is that most of the observed M31 CHVCs
are gas clouds in tidal streams, such as the Magellanic Stream, and
are not associated with distinct dark matter halos
\citep{putman_etal03}.

%--------------------------------------
\section{Comparison with previous work}
\label{sec:prevwork}
%--------------------------------------

Possible astrophysical solutions
\footnote{The solutions that invoke astrophysics of galaxy formation
within the standard CDM framework rather than modifications of the
properties of dark matter particles and/or the shape of the initial
power spectrum.}
to the ``missing satellite problem'' have been considered in the last
several years.  Here we discuss the main differences of our model and
the models proposed in previous studies.

\citet[][]{bullock_etal00}, \citet{somerville02}, and
\citet{benson_etal02} discussed the formation and evolution of dwarf
galactic satellites using semi-analytic models of different degrees of
sophistication. The conclusion reached by all these studies is that
the extragalactic UV background can greatly suppress the gas accretion
and star formation in the majority of low-mass ($V_{\rm m}\lesssim
30\ \rm km\,s^{-1}$) halos. A small fraction of the dwarf halos that
harbors stellar systems was assumed to have formed (i.e., assembled
significant fraction of their mass) before reionization, when the
level of UV radiation was low.  This is because in all of these
studies the maximum circular velocity of subhalos was assumed to be
constant as the mass is tidally stripped. There was thus a simple
one-to-one mapping between the circular velocity observed at $z=0$ and
at the time of accretion.  Our results show that this assumption is
incorrect \citep[see also][]{hayashi_etal03,kazantzidis_etal03}.
Another key difference is that tidal mass loss in our model can occur
before a halo is accreted by the host, as a result of interactions
with other halos.  These effects are not accounted for in any of the
semi-analytic models.

The implicit assumption in the above models is that the small systems
would be able to retain the accreted gas and form stars after
reionization.  This assumption was justified at the time, as the first
calculations of photoevaporation of gas indicated that halos with
$V_{\rm m}\gtrsim 10\rm\ km\,s^{-1}$ might retain their gas
\citep{barkana_loeb99}.  More recent calculations, however, show that
the gas could be gradually removed from halos of up to $V_{\rm
max}\approx 30\rm\ km\,s^{-1}$ \citep{shaviv_dekel03}. In light of
this result, the previous models would not be able to explain the
formation and properties of luminous dwarfs, as the star formation in
small halos would be suppressed after reionization.  It would thus be
difficult to explain the more extended star formation histories
derived for many dSph galaxies in the Local Group \citep{grebel00}.

In our model, the small-mass dwarfs are identified with the halos that
were relatively massive at high redshift and could retain the gas and
form stars after reionization.  The star formation histories of dwarfs
are thus more extended, in better accord with observations. As noted
in the previous section, our model is also insensitive to the epoch of
reionization and can accommodate early reionization suggested by
results of the {\sl WMAP} satellite \citep{spergel_etal03}.

Our model and all of the models discussed above are qualitatively
different from the proposal of \citet{stoehr_etal02,stoehr_etal03}.
These authors argued that the maximum circular velocity of the Local
Group dwarfs may be systematically underestimated because it is
derived from the stellar velocity dispersion within radii considerably
smaller than $r_{\rm max}$, the radius at which the maximum halo
velocity, $V_{\rm m}$ is reached \citep[see,
however,][]{kazantzidis_etal03}.  \citet[][see also \citet{hayashi_etal03}]{stoehr_etal02} suggested that
the luminous dwarfs may be harbored by the most massive satellites of
the DM halos.  This has an important physical implication: if the
dwarfs indeed occupy twelve or so most massive halos, then there
exists a certain mass scale below which galaxy formation is completely
suppressed.  If, on the other hand, the dwarf galaxies occupy
satellites with a variety of masses ($\sim 10^7-10^{10}\ \rm
M_{\odot}$), one has to explain why some fraction of small halos
managed to light up the stars, while most others did not.

If the idea of \citet{stoehr_etal02} is correct, our results indicate
that circular velocities of dwarf spheroidal halos should have been
even larger (by a factor of two or more) than the values inferred from
the current observations.  This could make halos of some galaxies
uncomfortably massive.  For example, \citet{stoehr_etal02} derive the
maximum circular velocity for the Draco in the range $\sim 35-55\rm\
km\,s^{-1}$.  This implies the pre-accretion values of $V_{\rm
max}\gtrsim 70\rm\ km\,s^{-1}$ and the pre-accretion mass comparable
to those of M32, NGC 205, and M33.  The fact that luminosity of Draco
is almost four orders of magnitude lower than luminosities of these
galaxies would present a major puzzle.

In addition, the radial distribution of the most massive satellites
should be consistent with the observed radial distribution of the MW
satellites.  We find that in our simulations the radial distribution
of subhalos with largest $V_{\rm m}$ is between that of the luminous
satellites and all DM satellites shown in Figure~\ref{fig:rdlg}.
 In a study of a larger sample of cluster halos,
\citet{delucia_etal03} find that the radial distribution of the most
massive halos is even more extended than that of the smaller mass
objects.  A similar point was made recently by \citet{taylor_etal03},
who used semi-analytic models for subhalo population to show that the
radial distribution of the most massive halos is more extended than
that of the MW satellites at $\gtrsim 3\sigma$ level.
A caveat to this argument is that the sample of Milky
Way satellites may be incomplete at large distances and more faint 
dwarf galaxies will be discovered in the future \citep{willman_etal04}.

%---------------------
\section{Conclusions}
\label{sec:conclusions}
%--------------------

We presented a study of the dynamical evolution of galactic satellites
using self-consistent high-resolution cosmological simulations of
three MW-sized halos. Our main results and conclusions are as follows.

\begin{itemize}
\item We find that $\approx 10\%$ of the substructure halos
that have masses of $<10^8-10^9\rm\ M_{\odot}$ at the present epoch,
had considerably higher masses and circular velocities when they
formed at $z>2$.  After the initial period of mass accretion, while
these objects evolve in isolation, they suffer dramatic mass loss due
to tidal stripping by actively merging massive neighboring
halos. Strong tidal interactions can occur even before the dwarfs are
accreted by their primary host halos.

\item The decrease in mass due to tidal stripping is accompanied by
the decrease in the maximum circular velocity, such that the objects
evolve along a $M-V_{\rm m}^{\alpha}$ relation with $\alpha \approx
3-4$.

\item These results indicate that some of the systems that have small
masses and circular velocities at $z=0$ could have had masses
comparable to those of the SMC and LMC in the past. This can explain
how the smallest dwarf spheroidal galaxies observed in the Local Group
were able to build up sizable stellar masses in such shallow potential
wells.

\item We present a simple galaxy formation model based on the
  evolutionary tracks extracted from the simulations.  The novel
  features of the model are the starburst mode of star formation after
  the strong peaks of the tidal force and accounting for the
  inefficient dissipation of gas in halos with $T_{\rm vir} \lesssim
  10^4$~K.  The model can successfully reproduce the circular velocity
  function, radial distribution, morphological segregation of the
  observed Milky Way satellites, and the basic properties of galactic
  dwarfs such as stellar masses and densities.

\end{itemize}

\acknowledgements 

We would like to thank Nick Gnedin for providing us with the results
of his filtering mass calculation in numerical form and for his
hospitality at the University of Colorado at Boulder where this paper
was completed.  We are also grateful to Andrew Zentner for useful
comments and Stelios Kazantzidis for communicating results of his
calculations prior to publication.  The simulations presented here
were performed on the Origin2000 at the National Center for
Supercomputing Applications (NCSA).  This work was supported by the
National Science Foundation under grant No.  AST-0206216 and
AST-0239759 to the University of Chicago.  OYG is supported by the
STScI Fellowship.

\bibliography{ms}

\appendix

%----------------------------
\section{Calculation of the tidal force}
\label{sec:tidalmodel}
%----------------------------

In our analysis we use the external tidal force experienced by each
satellite halo to estimate the strength of tidal interaction.  We calculate
the force both directly from the gravitational potential computed in
the simulation and using an analytical approximation for the
neighbor halos.

To compute the tidal force numerically from the local potential
$\Phi$, we estimate its second spatial derivative at the
center-of-mass of the satellite:
\begin{equation}
F_{\alpha} \equiv 
  - \left( {d^2\Phi \over dR_\alpha dR_\beta} \right)_0 \, r_\beta
  \equiv F_{\alpha\beta} \, r_\beta,
\end{equation}
where $\rr$ is the radius-vector in the satellite reference frame and
$\RR$ is the radius-vector in the perturber reference frame. The
potential $\Phi$ is calculated on the original refinement grid using
the ART gravity solver.  In the calculation of the potential,
we subtract the self contribution of the halo and consider only the
external tidal potential.

In a study of galaxy interactions in clusters of galaxies,
\citet{gnedin03a} used the Savitzky-Golay smoothing filter to
interpolate the potential on a plane and calculate its derivatives
from a smooth polynomial function.  We employ a similar scheme but
with the adaptive 4-th order interpolating polynomials in each of the
three orthogonal planes around the satellite center of mass:
\begin{equation}
  P_4(x,y) = \sum_{k,l=0}^4 c_{kl} x^k y^l
  \label{eq:p4}
\end{equation}
and the same for the $xz$ and $yz$ planes.  The 4-th order expansion
ensures a smooth second derivative of the potential.  In each of the
planes we extract a $n \times n$ subgrid centered on the original grid
point, nearest to the satellite center.  In order to obtain a uniform
accuracy of the tidal force for satellites of different sizes, we
choose the size of the subgrid cells to be closest to 1/4 of the
satellite's tidal radius.  The coefficients $c_{kl}$ are calculated by
minimizing $\chi^2$ deviation
\begin{equation}
  \chi^2 = \sum_{i,j=1}^n \left[{ P_4(x_i,y_j) - \Phi(x_i,y_j) }\right]^2
\end{equation}
using the CERN Program Library routine MINUIT\footnote{\tt
http://wwwasdoc.web.cern.ch/wwwasdoc/minuit/}.  We have experimented
with $n=16$, 32, and 64 and found that $n=64$ provides the most
accurate derivatives, as tested on the analytical NFW models.  The
tidal tensor components $F_{\alpha\beta}$ are then calculated by
analytical differentiation of equation (\ref{eq:p4}).

We compare the real tidal force due to the overall mass distribution
in the simulation with the contributions of all neighboring halos,
including the host halo.  We model the halos with an NFW density
profile and take their mass $M_{\rm vir}$ and virial radius $r_{\rm
vir}$ directly from the halo catalogs generated by the halo finder
(see \S~\ref{sec:haloid}).  We determine the scale radius of the NFW
model for the satellite halos from the position of the maximum
circular velocity, $r_s = r_{\rm max}/2.16$.  For the host halo, we use
the parametrization $c_{\rm nfw} \equiv r_{\rm vir}/r_s = 16 \,
a^{3/2}$, which is a best fit to the density profile of the analyzed
host halos.  The analytical tidal force in the reference frame of the
satellite is then readily calculated using eq. (5) of
\cite{gnedin_etal99}:
\begin{equation}
  {\bf F}(\rr) = {G M(R) \over R^3} 
       \left[ (3-\acute{\mu})({\bf n}\cdot\rr) - \rr \right]
\label{eq:Ftidnfw}
\end{equation}
where $\rr$ is the radius-vector within the satellite, $R$ is the
distance to the perturber, ${\bf n} \equiv \RR/R$, $\acute{\mu}
\equiv d\ln{M}/d\ln{R}$, and $M(R)$ is the enclosed mass of the
NFW model:
\begin{equation}
  M(R) = M_{\rm vir} \ {\ln{(1+R/r_s)} - 1 + (1 + R/r_s)^{-1} \over
                        \ln{(1+c_{\rm nfw})} - 1 + (1 + c_{\rm nfw})^{-1}}.
\end{equation}
Figure~\ref{fig:tr1} shows that the approximate tidal force calculated
in this manner is quite accurate, especially near the maximum of the
tidal force.

Although the tidal force along the satellite trajectory varies rapidly
with time, most of the tidal heating of stars and dark matter
particles occurs near the strong peaks of the tidal force.  Each of
these tidal peaks can be considered as an independent tidal shock
\citep{gnedin_ostriker99,gnedin03a}.  The amount of tidal heating,
such as the increase of the velocity dispersion, is proportional to
the integral over the peak of tidal force:
\begin{equation}
I_{\rm tid}(t_n) \equiv \sum_{\alpha,\beta} 
  \left( \int F_{\alpha\beta} \, dt \right)_n^2 \, 
  \left( 1 + {\tau_n^2 \over t_{\rm dyn}^2} \right)^{-3/2},
  \label{eq:Itid}
\end{equation}
where the sum extends over all components of the tidal tensor,
$\alpha,\beta=\{x,y,z\}$.  The last factor is the correction for the
conservation of adiabatic invariants of stellar orbits during the
tidal shock \citep[c.f.,][]{gnedin_ostriker99}.  Here $\tau_n$ is the
effective duration of peak $n$ at time $t_n$, and $t_{\rm dyn}$ is the
dynamical time of the satellite.  We take $t_{\rm dyn} = 2\pi
r_{1/2}/v_{\rm rot}$, where $r_{1/2}$ is the half-mass radius of the
stellar disk and $v_{\rm rot}$ is the circular velocity of the
appropriate NFW model at $r_{1/2}$.  The cumulative tidal heating
parameter is the sum over all tidal peaks:
\begin{equation}
I_{\rm tid} = \sum_n I_{\rm tid}(t_n).
  \label{eq:Itidsum}
\end{equation}
This parameter determines the increase of the velocity
dispersion of stars (eq. [\ref{eq:sigma}]) in our model of dwarf
galaxy formation (\S\ref{sec:sam}).

%----------------------------------------------
\section{Filtering mass scale}
\label{sec:filter}
%----------------------------------------------

We estimate the suppression of gas accretion due to the extragalactic
UV background using the filtering mass scale, derived by
\citet{gnedin00}.  He defined $M_{\rm F}$ as the mass of the halo
which would lose half of the baryons, compared to the universal baryon
fraction.  This {\em filtering mass} relates to the Jeans mass of the
intergalactic gas integrated over the cosmic history (eq. [6] in
\citealt{gnedin00}):
\begin{eqnarray}
  M_{\rm F}(a) & = & M_{\rm J0} \ f(a)^{3/2}, \label{eq:mf}\\
  f(a) & = & {3 \over a} \int_0^a x \, T_4(x)
             \left[{1 - \left({x \over a}\right)^{1/2}}\right] dx \nonumber
\end{eqnarray}
where $M_{\rm J0} = 2.5\times 10^{11} h^{-1} \Omega_0^{-1/2} \mu^{-3/2} \
M_{\sun}$, $\mu \approx 0.59$ is the mean molecular weight of the
fully ionized gas, and the integration extends over the expansion
factor, $a$.  The temperature of the cosmic gas $T_4$ is expressed in
units of $10^4$ K for convenience.

\begin{figure}[t]
\vspace{-1cm}
\centerline{\epsfysize3.5truein \epsffile{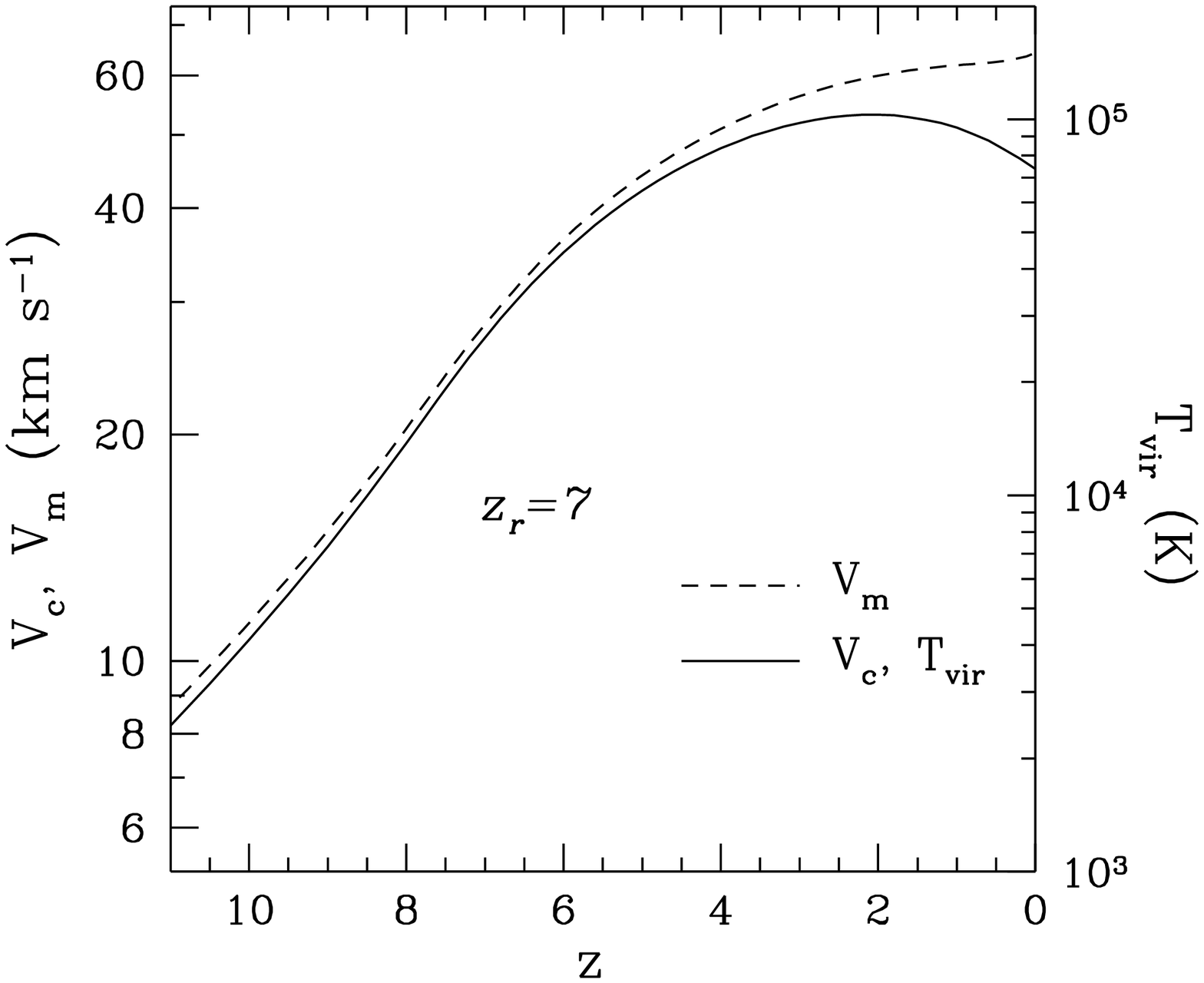}}
\vspace{-1cm}
\centerline{\epsfysize3.5truein \epsffile{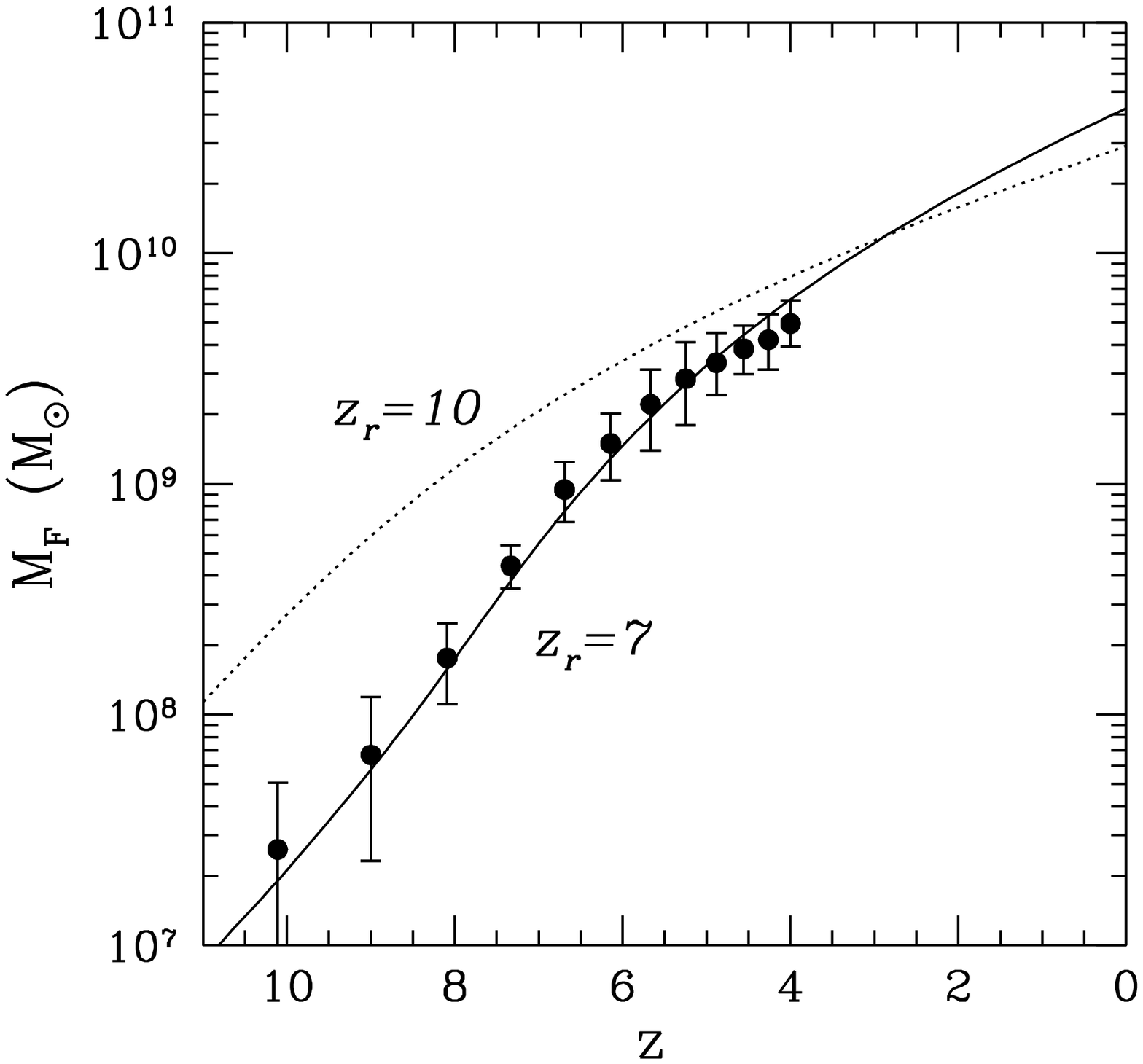}}

\caption{{\it Top panel:} filtering virial circular velocity ($V_{\rm
c}$) and virial temperature (axes are chosen such that the two solid
lines coincide), and maximum circular velocity ($V_{\rm m}$, dashed
line).  {\it Bottom panel:} filtering mass.  Data points with
error-bars show the simulation results of \citet{gnedin00}.  Solid
lines are for the standard epoch of reionization: $z_r = 7$, $z_o =
8$.  For comparison, the dotted line in the bottom panel shows the case of
earlier reionization with $z_r = 10$, $z_o = 11$.
\label{fig:vmfilt}}
\end{figure}

Here we propose an analytical fit to the results of \citet{gnedin00},
assuming a simple dependence of the temperature on the expansion
factor: $T_4(a) = (a/a_o)^{\alpha}$ for $a \le a_o$, $T_4(a) = 1$ for
$a_o \le a \le a_r$, and $T_4(a) = (a/a_r)^{-1}$ for $a \ge a_r$.

These three distinct stages can be clearly seen on Fig. 1 of
\citet{gnedin00}. They correspond, to the 1) epoch before the first
HII regions form, $z>z_o$, 2) the epoch of the overlap of multiple HII
regions, $z_r<z<z_o$, and 3) the epoch of complete reionization,
$z<z_r$.  In the first stage, before redshift $z_o \equiv 1/a_o-1
\approx 8$, the temperature is rising as the newly-formed stars ionize
their neighboring regions.  The parameter $\alpha$ controls the rate
of growth of the extragalactic UV flux; we find $\alpha = 6$ to be the
best fit.  During the overlap stage, between redshifts $z_o$ and $z_r
\equiv 1/a_r-1 \approx 7$, the temperature is kept constant at roughly
$10^4$ K as the cosmic HII regions overlap.  After the universe is
fully ionized, at redshifts below $z_r$, the temperature falls
adiabatically with the cosmic expansion.

With these analytical expressions for $T_4(a)$, we integrate equation
(\ref{eq:mf}) analytically:
\begin{eqnarray}
f(a) &=& {3a \over (2+\alpha)(5+2\alpha)} \left({a \over a_o}\right)^\alpha,
  \quad a \le a_o \\
f(a) &=& {3 \over a} \left\{ a_o^2 
          \left[{{1 \over 2+\alpha} - {2(a/a_o)^{-1/2} \over 5+2\alpha}}\right]
          \right. \nonumber\\
     && \left. + {a^2 \over 10} - {a_o^2 \over 10} 
          \left[{ 5 - 4 \left({a/a_o}\right)^{-1/2}}\right] \right\},
  \quad a_o \le a \le a_r \nonumber\\
f(a) &=& {3 \over a} \left\{ a_o^2
          \left[{{1 \over 2+\alpha} - {2(a/a_o)^{-1/2} \over 5+2\alpha}}\right]
          \right. \nonumber\\
     &&  + {a_r^2 \over 10} \left[{ 5 - 4 \left({a/a_r}\right)^{-1/2}}\right]
         - {a_o^2 \over 10} \left[{ 5 - 4 \left({a/a_o}\right)^{-1/2}}\right]
          \nonumber\\
     &&  \left. + {a a_r \over 3} 
          - {a_r^2 \over 3} \left[{ 3 - 2 (a/a_r)^{-1/2}}\right] \right\},
  \quad a \ge a_r. \nonumber
\end{eqnarray}

The virial circular velocity of the halo is \\ $V_c^3 = G M H(z)
(\Delta_{\rm vir}/2)^{1/2}$, where $H(z) = H(0) [\Omega_0 (1+z)^3 +
\Omega_\Lambda]^{1/2}$ is the Hubble constant, and $\Delta_{\rm vir}$
is the virial overdensity with respect to the critical density,
parametrized by \citet{bryan_norman98} as $\Delta_{\rm vir}(z) =
18\pi^2 + 82 x - 39 x^2$, $x \equiv \Omega(z)-1$.  The virial
temperature is $T_{\rm vir} = 36 (V_c / \mbox{km s}^{-1})^2$ K.

Our analytical fit is convenient for accurate modeling of the
photoheating effect in semi-analytical models of galaxy formation.
Its versatile form, with two parameters $z_o$ and $z_r$, allows a
simple recalculation of the filtering mass for a different redshift of
reionization than was assumed in the simulation of \citet{gnedin00}.
It can also be easily adapted to describe two epochs of reionization,
or the early extended reionization suggested by WMAP
\citep{spergel_etal03}.  For illustration, we show on the lower panel
of Figure~\ref{fig:vmfilt} the filtering mass as a function of
redshift for two choices of the reionization redshift.  The top panel
shows the filtering circular velocity and the corresponding values of
the maximum velocity $V_{\rm m}$ and virial temperature $T_{\rm vir}$.

\citet{gnedin00} provided the following expression for the amount of
cold gas left in the halo of mass $M$:
\begin{equation}
  f_{\rm g}(M,z) = {f_{\rm b} \over [1 + 0.26 M_{\rm F}(z)/M]^3},
  \label{eq:fg}
\end{equation}
where $f_{\rm b} \approx 0.14$ is the universal baryon fraction.
In \S~\ref{sec:sam} we use this fraction of cold gas to model the star
formation history of the satellite galaxies.  To account for the
inefficiency of atomic gas cooling at $T < 10^4$ K, we apply equation
(\ref{eq:fg}) substituting for $M_{\rm F}$ the maximum of $M_{\rm F}(z)$ 
or $M_4$, the halo mass corresponding to $T_{\rm vir} = 10^4$ K.

\end{document}